# AN INVERSE PROBLEM TO DETERMINE THE SHAPE OF A HUMAN VOCAL TRACT

Tuncay Aktosun
Department of Mathematics
University of Texas at Arlington
Arlington, TX 76019-0408, USA

Paul Sacks
Department of Mathematics
Iowa State University
Ames, IA 50011, USA

Xiao-Chuan Xu
School of Mathematics and Statistics
Nanjing University of Information Science and Technology
Nanjing, 210044, Jiangsu, People's Republic of China

**Abstract**: The inverse problem of determining the cross-sectional area of a human vocal tract during the utterance of a vowel is considered. The frequency-dependent boundary condition at the lips is expressed in terms of the acoustic impedance of a vibrating piston on an infinite plane baffle. The corresponding pressure at the lips is expressed in terms of the normalized impedance and a key quantity related to the so-called Jost function of an associated Schrödinger equation. Various input data sets are considered to solve the inverse problem, including the pressure at the lips, the absolute pressure at the lips, or the poles of the pressure at the lips. The solution to the inverse problem is obtained with the help of the Gel'fand-Levitan method for the associated Schrödinger equation.

# 1. INTRODUCTION

The human speech consists of units called phonemes. For example, in uttering the word "book," the phonemes /b/, /u/, and /k/ are produced in succession typically at the rate of 10 phonemes per second. The phonemes can be classified into two main groups, i.e. into vowels and consonants. To a good approximation the production of each vowel can be assumed to occur in the vocal tract, by ignoring the articulators such as the nasal cavity and the tongue. During the utterance of each phoneme it can be assumed that the vocal tract is a circular cylindrical tube of some fixed length $\ell$, which is about 17 cm for an adult male.

The vocal tract can be parameterized by using the independent variable $x$, which denotes the distance from the glottis (the opening between the vocal cords). The radius of the vocal tract at location $x$ can be denoted by $r(x)$, with the point $x = 0$ corresponding to the glottis and the point $x = \ell$ corresponding to the lips. For a mathematical description we can assume that in the vowel production the radius function belongs to the class $\mathcal{A}$ defined below, where we recall that $\ell$ is a fixed constant. Throughout the paper we use a prime to denote the derivative with respect to the independent spatial variable.

**Definition 1.1** *The vocal-tract radius $r$ belongs to class $\mathcal{A}$ if the following conditions are satisfied:*

(a) *The quantity $r(x)$ is real valued and positive for $x \in [0, \ell]$, where we let*

$$r_0 := r(0), \quad r_\ell := r(\ell). \tag{1.1}$$

(b) *The derivative function $r'$ is continuous for $x \in [0, \ell]$, where we let*

$$r'_0 := r'(0), \quad r'_\ell := r'(\ell). \tag{1.2}$$

*We remark that $r'_0$ and $r'_\ell$ may be zero, positive, or negative.*

(c) *The second-derivative function $r''$ is integrable on $x \in (0, \ell)$.*

The production of each phoneme in the vocal tract can be described by specifying the sound pressure $p(x,t)$ and the volume velocity $v(x,t)$ at location $x$ and at time $t$.

Equivalently, we can use the Fourier transformation from the time domain to the frequency domain as

$$P(k,x) = \int_{-\infty}^{\infty} dt\, p(x,t)\, e^{-ikct}, \quad V(k,x) = \int_{-\infty}^{\infty} dt\, v(x,t)\, e^{-ikct}, \tag{1.3}$$

where $k$ is the angular wavenumber measured in radians per second and $c$ is the sound speed in the vocal tract, which can be assumed to have the constant value $3.5 \times 10^4$ cm/sec in the vocal tract at the normal body temperature $37°$C. Since the quantities $p(x,t)$ and $v(x,t)$ are real valued, from (1.3) it follows that

$$P(-k,x) = P(k,x)^*, \quad V(-k,x) = V(k,x)^*, \qquad k \in \mathbf{R}, \quad x \in [0,\ell], \tag{1.4}$$

where we use $\mathbf{R}$ to denote the real line and the asterisk denotes complex conjugation.

The angular wavenumber $k$ and the frequency $\nu$ are related to each other as

$$k = \frac{2\pi\nu}{c}.$$

Since the sound speed $c$ in the vocal tract is a constant, we can view the angular wavenumber as a measure of the frequency with the proportionality constant $2\pi/c$, and hence we have $k = 1.8 \times 10^{-4} \nu$ with $k$ expressed in rad/cm and $\nu$ expressed in Hertz (cycles per second). The air density $\mu$ also affects the sound production in the vocal tract, and its value is about $1.14 \times 10^{-3}$ gm/cm$^3$ in the vocal tract at the normal body temperature.

The production of each phoneme in the vocal tract is governed by the first-order system of differential equations [1,11,12,17,18,20-23]

$$\begin{cases} \pi r(x)^2\, P'(k,x) + ikc\mu\, V(k,x) = 0, \\ c\mu\, V'(k,x) + ik\,\pi\, r(x)^2\, P(k,x) = 0, \end{cases} \tag{1.5}$$

where the independent variable $x$ is confined to the interval $(0,\ell)$, we recall that the prime denotes the $x$-derivative, and $P(k,x)$ and $V(k,x)$ are the dependent variables for $k \in \mathbf{R}$. Eliminating one of the dependent variables, we can convert (1.5) into the second-order equation

$$[r(x)^2\, P'(k,x)]' + k^2\, r(x)^2\, P(k,x) = 0, \qquad x \in (0,\ell), \tag{1.6}$$

or into
$$\left[\frac{V'(k,x)}{r(x)^2}\right]' + k^2 \frac{V(k,x)}{r(x)^2} = 0, \qquad x \in (0,\ell). \tag{1.7}$$

We remark that (1.6) was first derived by Webster [23] and is known as the Webster horn equation.

In order to determine $P(k,x)$ and $V(k,x)$ appearing in (1.5) uniquely, we can provide some appropriate physical restrictions at $x = 0$ and $x = \ell$. An appropriate condition commonly used at $x = 0$ is

$$v(0,t) = \delta(t), \tag{1.8}$$

which amounts to choosing the volume velocity $v(x,t)$ at the glottis as the Dirac delta distribution. From (1.3) and (1.8) it follows that

$$V(k,0) \equiv 1, \qquad k \in \mathbf{R}, \tag{1.9}$$

or equivalently, with the help of the first line of (1.5) we have

$$P'(k,0) = -\frac{ikc\mu}{\pi r_0^2}. \tag{1.10}$$

A reasonable physical restriction at $x = \ell$ is obtained by assuming that the sound pressure at the lips goes out of the mouth and there is no component of the sound pressure moving into the mouth at $x = \ell$. Mathematically, this amounts to the the boundary condition at $x = \ell$ given by

$$P'(k,\ell) + \left[ik + \frac{r'_\ell}{r_\ell}\right] P(k,\ell) = 0. \tag{1.11}$$

For further information on the boundary condition specified in (1.11) we refer the reader to [3-5] and the references therein.

The boundary condition given in (1.11) is not the only boundary condition that can be chosen at $x = \ell$. In mathematical studies of human speech, the scientists with electrical engineering background usually describe the human vocal tract as an acoustic tube with a piecewise constant radius function $r(x)$, viewed as a cascade of networks containing analogous circuit elements with $P(k,x)$ and $V(k,x)$ playing the roles of the electric voltage and the electric current, respectively. The impedance at the lips, which we denote by $Z(k)$, is defined as

$$Z(k) := \frac{P(k,\ell)}{V(k,\ell)}, \qquad k \in \mathbf{R}, \tag{1.12}$$

and the normalized impedance at the lips is defined as

$$z(k) := \frac{\pi r_\ell^2}{c\mu} Z(k), \qquad k \in \mathbf{R}. \tag{1.13}$$

By viewing each component of the electric circuit as a parallel connection of a resistor and an inductor, we can express the normalized impedance at the lips also as

$$\frac{1}{z(k)} = \frac{1}{R} + \frac{1}{ikcL},$$

where $R$ and $L$ denote the corresponding frequency-dependent normalized resistance and inductance, respectively.

If the impedance at the lips is known, then we can view (1.12) as a boundary condition at $x = \ell$, and with the help of the first line of (1.5), (1.12), and (1.13) we can express the corresponding boundary condition at $x = \ell$ as

$$P'(k,\ell) + \frac{ik}{z(k)} P(k,\ell) = 0, \qquad k \in \mathbf{R}. \tag{1.14}$$

Hence, specifying the normalized impedance at the lips is equivalent to having the boundary condition (1.14). We refer the reader to Section 1.1 of [12] for some appropriate choices for $z(k)$. We can view a typical human head as a sphere, where the radius of the sphere is about 9 cm for an adult male. When $r_\ell$ is small compared to the radius of the head, the normalized impedance at the lips can be approximated [12,17] in terms of the Bessel function of order one, i.e. $J_1(w)$, and the Struve function of order one, i.e. $\mathbf{H}_1(w)$, as

$$z(k) = 1 - \frac{J_1(2kr_\ell)}{kr_\ell} + i\frac{\mathbf{H}_1(2kr_\ell)}{kr_\ell}. \tag{1.15}$$

The choice of $z(k)$ can be described as the sound radiation at the lips being governed by a vibrating piston on an infinite plane baffle. Throughout the paper, when we refer to the determination of $z(k)$ we mean the determination of the specific form of $z(k)$ given in (1.15), which is obtained when $r_\ell$ is known.

Let us consider the following direct and inverse scattering problems for the production of a vowel in the human vocal tract. Our direct problem consists of the determination of the sound pressure $P(k, \ell)$ at the lips for $k \in \mathbf{R}$ when the vocal tract radius $r(x)$ is given for $x \in (0, \ell)$ and is known to belong to the class $\mathcal{A}$ described in Definition 1.1. Our inverse problem consists of the determination of $r(x)$ for $x \in (0, \ell)$ when $P(k, \ell)$ is known for $k \in \mathbf{R}$. A modified inverse problem consists of the determination of $r(x)$ for $x \in (0, \ell)$ when the absolute pressure $|P(k, \ell)|$ is known for $k \in \mathbf{R}$. Note that because of the first equality in (1.4) the pressure $P(k, x)$ known for $k \geq 0$ determines the pressure $P(k, x)$ for $k \in \mathbf{R}$. We refer the reader to [5] for the solutions to the corresponding direct and inverse scattering problems when the boundary conditions (1.10) at $x = 0$ and (1.11) at $x = \ell$ are used. In this paper we analyze the corresponding direct and inverse scattering problems by using the boundary conditions (1.10) at $x = 0$ and (1.14) at $x = \ell$ with $z(k)$ as in (1.15). A main contribution of this paper is the solution of the inverse problem with the boundary condition involving the special form of the impedance given in (1.15).

Our paper is organized as follows. In Section 2 we present the relevant properties of the normalized impedance $z(k)$, including its small-$k$ asymptotics and large-$k$ asymptotics. In Section 3 we explore the relationship between the Webster horn equation (1.6) and the corresponding Schrödinger equation (3.2) whose potential $q(x)$ is related to the radius function $r(x)$ as in (3.3). We introduce certain particular solutions to (3.2) and present their relevant properties, including their small-$k$ asymptotics and large-$k$ asymptotics. Two such solutions, the regular solution $\varphi(k, x)$ appearing in (3.6) and another regular solution $g(k, x)$ appearing in (3.7) are used to express the key quantity $G(k)$ defined in (3.15). In the rest of Section 3 and in Section 4 we present various relevant properties of $G(k)$. In Section 4 the zeros of $G(k)$ are analyzed and $G(k)$ is related to another key quantity, the so-called Jost function $F(k)$ appearing in (4.30). In Section 5 we present the solution $P(k, x)$ to the Webster horn equation (1.6) with the boundary conditions (1.10) and (1.14).

This enables us the express the pressure at the lips in terms of the normalized impedance $z(k)$ and the key quantity $G(k)$, as in (5.9). In Section 6 we present the solution to the inverse problem of recovery of $r(x)$ from some input data set related to $P(k, \ell)$. We identify such a data set and also present various equivalents, which are expressed in terms of the poles of $P(k, \ell)$, the absolute pressure $|P(k, \ell)|$, or $P(k, \ell)$ itself. Using the key relation (5.9) we show that $r(x)$ can be recovered by construction the Jost function $F(k)$ and using $F(k)$ as input to the Gel'fand-Levitan method arising in the inverse scattering theory for the Schrödinger equation. We clarify whether the vocal tract length $\ell$ should be a part of the input data set or it can be recovered from some input data set. Finally, in Section 7 we illustrate some aspects of the theory developed via some explicit examples. In particular, we illustrate the relationship between the density of the zeros of $G(k)$ for large $k$ and the constants $\ell$ and $r_\ell$, where $r_\ell$ is the radius of the opening between the lips.

## 2. THE PROPERTIES OF THE NORMALIZED IMPEDANCE

In this section we establish the relevant properties of the normalized impedance given in (1.15). In particular, we present the properties related to the analyticity and the small-$k$ and large-$k$ asymptotics of $z(k)$.

**Proposition 2.1** *Assume that $r_\ell > 0$, which is the case when the vocal-tract radius $r(x)$ belongs to the class $\mathcal{A}$ specified in Definition 1.1. Let $z(k)$ be the quantity given in (1.15) expressed in terms of $J_1(w)$ and $\mathbf{H}_1(w)$. Then, we have the following:*

(a) *The quantity $z(k)$ is entire in $k \in \mathbf{C}$, where we use $\mathbf{C}$ to denote the complex plane.*

(b) *We have the symmetry property*

$$z(-k^*) = z(k)^*, \qquad k \in \mathbf{C}. \tag{2.1}$$

(c) *The quantity $z(k)$ has a simple zero at $k = 0$, and its small-$k$ asymptotics of $z(k)$ is given by*

$$z(k) = \frac{8ikr_\ell}{3\pi} + \frac{k^2 r_\ell^2}{2} + O(k^3), \qquad k \to 0 \text{ in } \mathbf{C}. \tag{2.2}$$

(d) *When $k \in \mathbf{R}$ the real part of $z(k)$ given by*

$$\operatorname{Re}[z(k)] = 1 - \frac{J_1(2kr_\ell)}{kr_\ell}, \qquad k \in \mathbf{R}, \tag{2.3}$$

*has a double zero at $k = 0$, is positive when $k > 0$, has the limit equal to one when $k \to +\infty$, and is even in $k \in \mathbf{R}$.*

(e) *When $k \in \mathbf{R}$ the imaginary part of $z(k)$ given by*

$$\operatorname{Im}[z(k)] = \frac{\mathbf{H}_1(2kr_\ell)}{kr_\ell}, \qquad k \in \mathbf{R}, \tag{2.4}$$

*has a simple zero at $k = 0$, is positive when $k > 0$, has the limit equal to zero when $k \to +\infty$, and is odd in $k \in \mathbf{R}$.*

(f) *When $k$ is on the imaginary axis in $\mathbf{C}$, the quantity $z(k)$ is real valued and is given by*

$$z(i\kappa) = 1 - \frac{4}{\pi} \int_0^1 dt\, e^{2\kappa r_\ell t} \sqrt{1 - t^2}, \qquad \kappa \in \mathbf{R}. \tag{2.5}$$

*The quantity $z(i\kappa)$ is a decreasing function in $\kappa \in \mathbf{R}$, has a simple zero at $\kappa = 0$, is negative when $\kappa > 0$, is positive when $\kappa < 0$, $z(i\kappa) \to 1$ as $\kappa \to -\infty$, and $z(i\kappa) \to -\infty$ as $\kappa \to +\infty$.*

(g) *For any $k$-value in $\mathbf{C}$, the quantity $z(-k)$ is obtained from (1.15) by replacing $+i$ there with $-i$, i.e. we have*

$$z(-k) = 1 - \frac{J_1(2kr_\ell)}{kr_\ell} - i\,\frac{\mathbf{H}_1(2kr_\ell)}{kr_\ell}. \tag{2.6}$$

PROOF: It is known [2] that $J_1(w)$ and $\mathbf{H}_1(w)$ can be expressed as the respective infinite series given by

$$J_1(w) = \sum_{j=0}^{\infty} \frac{(-1)^j\, w^{2j+1}}{2^{2j+1}\, j!\,(j+1)!}, \qquad w \in \mathbf{C}, \tag{2.7}$$

$$\mathbf{H}_1(w) = \sum_{j=0}^{\infty} \frac{(-1)^j\, w^{2j+2}}{2^{2j+2}\, \Gamma(j+3/2)\, \Gamma(j+5/2)}, \qquad w \in \mathbf{C}, \tag{2.8}$$

where $\Gamma(\alpha)$ denotes the gamma function. We recall that $\Gamma(\alpha)$ has various useful properties such as

$$\Gamma(\alpha+1) = \alpha\,\Gamma(\alpha), \qquad \alpha > 0; \quad \Gamma(1/2) = \sqrt{\pi}.$$

Using the ratio tests on the series in (2.7) and (2.8), one can establish that $z(k)$ is entire, which proves (a). From (2.7) and (2.8) we see that (2.1) holds, establishing (b), and that $z(k)$ has the small-$k$ asymptotics described in (2.2), which proves (c). Since $J_1(w)$ and $\mathbf{H}_1(w)$ are real valued when $w \in \mathbf{R}$, from (1.15) we see that, for $k \in \mathbf{R}$, the real part of $z(k)$ is given by (2.3), and from the properties [2] of $J_1(w)$ we obtain the proof of (d). Similarly, for $k \in \mathbf{R}$, we see that the imaginary part of $z(k)$ is given by (2.4), and using the properties [1,2] of $\mathbf{H}_1(w)$ we establish the proof of (e). The proof of (f) can be given as follows. From p. 360 and p. 498, respectively, in [2] we have

$$J_1(iy) = \frac{2iy}{\pi} \int_0^{\pi/2} d\theta \, \sin^2 \theta \, \cos(iy \cos \theta), \qquad y \in \mathbf{R},$$

$$\mathbf{H}_1(iy) = -\frac{2y}{\pi} \int_0^{\pi/2} d\theta \, \sin^2 \theta \, \sinh(y \cos \theta), \qquad y \in \mathbf{R},$$

which together yield

$$J_1(iy) - i\,\mathbf{H}_1(iy) = \frac{2iy}{\pi} \int_0^{\pi/2} d\theta \, \sin^2 \theta \, e^{2y \cos \theta}, \qquad y \in \mathbf{R}. \tag{2.9}$$

Using (2.9) in (1.15) we obtain

$$z(i\kappa) = 1 - \frac{4}{\pi} \int_0^{\pi/2} d\theta \, \sin^2 \theta \, e^{2\kappa r_\ell \cos \theta}, \qquad \kappa \in \mathbf{R},$$

which is equivalent to (2.5). From (2.5) we get

$$\frac{dz(i\kappa)}{d\kappa} = -\frac{8r_\ell}{\pi} \int_0^1 dt \, e^{2\kappa r_\ell t} \, t\sqrt{1-t^2}, \qquad \kappa \in \mathbf{R},$$

from which we see that the right-hand side is negative for $\kappa \in \mathbf{R}$. Thus, the real-valued quantity $z(i\kappa)$ is a decreasing function of $\kappa \in \mathbf{R}$. From (2.2) we already know that $z(i\kappa)$ has a simple zero at $\kappa = 0$, and hence we conclude that $z(i\kappa) > 0$ for $\kappa < 0$ and we have $z(i\kappa) < 0$ for $\kappa > 0$. Thus, the proof of (f) is complete. We obtain the proof of (g) as follows. From (2.7) we see that $J_1(2w)/w$ is even in $w \in \mathbf{C}$ and from (2.8) we observe that $\mathbf{H}_1(2w)/w$ is odd in $w \in \mathbf{C}$. Thus, replacing $k$ by $-k$ in (1.15) we obtain (2.6). Hence, the proof of (g) is complete. ∎

In the next proposition we establish the large-$k$ asymptotics of $z(k)$. We let $\mathbf{C}^+$ denote the upper-half complex plane, $\mathbf{C}^-$ the lower-half complex plane, $\overline{\mathbf{C}^+} := \mathbf{C}^+ \cup \mathbf{R}$, and $\overline{\mathbf{C}^-} := \mathbf{C}^- \cup \mathbf{R}$. We also use the notation $\mathbf{R}^+ := (0, +\infty)$ and $\mathbf{R}^- := (-\infty, 0)$.

**Proposition 2.2** *Assume that $r_\ell > 0$, which is the case when the vocal-tract radius $r(x)$ belongs to the class $\mathcal{A}$ specified in Definition 1.1. Let $z(k)$ be the quantity given in (1.15) expressed in terms of $J_1(w)$ and $\mathbf{H}_1(w)$. Then, the large-$k$ asymptotics of $z(k)$ in $\mathbf{C}$ is given by*

$$z(k) = 1 + \frac{1}{kr_\ell}\left[\frac{2i}{\pi} + B(kr_\ell)\left(1 + O\left(\frac{1}{k}\right)\right) + O\left(\frac{1}{k^2}\right)\right], \qquad k \to \infty \text{ in } \mathbf{C}, \qquad (2.10)$$

*where we have defined*

$$B(kr_\ell) := \frac{(1-i)\,e^{-2ikr_\ell}}{\sqrt{2\pi kr_\ell}}, \qquad (2.11)$$

*with the square-root function being the principal branch of the complex square-root function. Consequently, we have*

$$e^{2ikr_\ell} z(k) = \frac{(1-i)}{kr_\ell \sqrt{2\pi kr_\ell}}\left[1 + O\left(\frac{1}{k}\right)\right] + O\left(e^{2ikr_\ell}\right), \qquad k \to \infty \text{ in } \mathbf{C}^+, \qquad (2.12)$$

$$z(k) = 1 + \frac{2i}{\pi kr_\ell} + O\left(\frac{1}{|k|^{3/2}}\right), \qquad k \to \infty \text{ in } \overline{\mathbf{C}^-}. \qquad (2.13)$$

PROOF: From p. 368 of [24] we know that we can express $J_1(w)$ in terms of the Whittaker function $W_{0,1}(w)$ as

$$J_1(w) = \frac{1}{\sqrt{2\pi w}}\left[e^{3\pi i/4} W_{0,1}(2iw) + e^{-3\pi i/4} W_{0,1}(-2iw)\right]. \qquad (2.14)$$

On the other hand, from p. 497 of [2] we have

$$\mathbf{H}_1(w) = \frac{2}{\pi} + Y_1(w) + O\left(\frac{1}{w^2}\right), \qquad w \to \infty \text{ in } \mathbf{C} \setminus \mathbf{R}^-. \qquad (2.15)$$

where $Y_1(w)$ is the Bessel function of the second kind, which can be expressed in terms of the Whittaker function $W_{0,1}(w)$ as

$$Y_1(w) = \frac{1}{\sqrt{2\pi w}}\left[e^{5\pi i/4} W_{0,1}(2iw) + e^{-5\pi i/4} W_{0,1}(-2iw)\right]. \qquad (2.16)$$

Using
$$e^{3\pi i/4} - i\,e^{5\pi i/4} = -\sqrt{2}\,(1-i), \quad e^{-3\pi i/4} - i\,e^{-5\pi i/4} = 0,$$

from (2.14)-(2.16) we get

$$J_1(w) - i\,\mathbf{H}_1(w) = -\frac{2i}{\pi} - \frac{1-i}{\sqrt{\pi w}}\,W_{0,1}(w) + O\left(\frac{1}{w^2}\right), \qquad w \to \infty \text{ in } \mathbf{C} \setminus \mathbf{R}^-. \tag{2.17}$$

From p. 343 of [2] we have

$$W_{0,1}(w) = e^{-w/2}\left[1 + O\left(\frac{1}{w}\right)\right], \qquad w \to \infty \text{ in } \mathbf{C}. \tag{2.18}$$

Using (2.18) in (2.17), as $w \to \infty$ in $\mathbf{C} \setminus \mathbf{R}^-$, we get

$$J_1(w) - i\,\mathbf{H}_1(w) = -\frac{2i}{\pi} - \frac{1-i}{\sqrt{\pi w}}\,e^{-w/2}\left[1 + O\left(\frac{1}{w}\right)\right] + O\left(\frac{1}{w^2}\right). \tag{2.19}$$

Finally using (2.19) in (1.15), we obtain (2.10) as $w \to \infty$ in $\mathbf{C} \setminus \mathbf{R}^-$. On the other hand, $z(k)$ is entire and satisfies (2.1) and hence (2.10) holds for $k \to \infty$ in $\mathbf{C}$. From (2.10) and (2.11), by retaining the leading terms, we obtain (2.12) as $k \to \infty$ in $\mathbf{C}^+$ and (2.13) as $k \to \infty$ in $\overline{\mathbf{C}^-}$, respectively. ∎

## 3. THE TRANSFORMATION TO THE SCHRÖDINGER EQUATION

In order to analyze the aforementioned direct and inverse problems related to the Webster horn equation (1.6), we can transform (1.6) into the more familiar Schrödinger equation and solve the direct and inverse problems for (1.6) in terms of the quantities related to the Schrödinger equation.

Letting
$$\psi(k,x) = r(x)\,P(k,x), \tag{3.1}$$

and using the fact that $P(k,x)$ is a solution to (1.6), we see that $\psi(k,x)$ satisfies the Schrödinger equation

$$-\psi''(k,x) + q(x)\,\psi(k,x) = k^2\,\psi(k,x), \qquad x \in (0,\ell), \tag{3.2}$$

where the potential $q(x)$ is related to the radius $r(x)$ of the vocal tract as

$$q(x) := \frac{r''(x)}{r(x)}, \qquad x \in (0, \ell). \tag{3.3}$$

There are various useful particular solutions to (3.2). One of them is the Jost solution $f(k, x)$ satisfying the initial conditions

$$f(k, \ell) = e^{ik\ell}, \quad f'(k, \ell) = ik\, e^{ik\ell}. \tag{3.4}$$

Another particular solution, which we denote by $S(k, x)$, to (3.2) is the sine-like solution satisfying the initial conditions

$$S(k, 0) = 0, \quad S'(k, 0) = 1. \tag{3.5}$$

The regular solution to (3.2), which we denote by $\varphi(k, x)$, satisfies the initial conditions

$$\varphi(k, 0) = 1, \quad \varphi'(k, 0) = \frac{r'_0}{r_0}, \tag{3.6}$$

where we recall that $r_0$ and $r'_0$ are the constants appearing in (1.1) and (1.2), respectively. We also introduce another particular solution, which we denote by $g(k, x)$, to (3.2) satisfying the initial conditions

$$g(k, \ell) = z(k), \quad g'(k, \ell) = \frac{r'_\ell}{r_\ell} z(k) - ik, \tag{3.7}$$

where $z(k)$ is the normalized impedance given in (1.15) and the quantities $r_\ell$ and $r'_\ell$ are the constants appearing in (1.1) and (1.2).

In the proposition below, we summarize the basic relevant properties of the aforementioned particular solutions to (3.2). These results will be useful in the analysis of direct and inverse problems for (1.6).

**Proposition 3.1** *Assume that the vocal-tract radius $r(x)$ belongs to the class $\mathcal{A}$ specified in Definition 1.1. Then, we have the following:*

(a) *The potential $q(x)$ defined in (3.3) is real valued and integrable for $x \in (0, \ell)$.*

(b) *Each of the particular solutions $f(k,x)$, $S(k,x)$, $\varphi(k,x)$, and $g(k,x)$ to (3.2) satisfying the initial conditions given in (3.4), (3.5), (3.6), and (3.7), respectively, exist and is unique for each $k \in \mathbf{C}$.*

(c) *For each fixed $x \in [0,\ell]$, the particular solutions $f(k,x)$, $S(k,x)$, $\varphi(k,x)$, and $g(k,x)$ and their $x$-derivatives $f'(k,x)$, $S'(k,x)$, $\varphi'(k,x)$, and $g'(k,x)$ are entire in $k \in \mathbf{C}$.*

(d) *Each of the particular solutions $f(k,x)$, $S(k,x)$, $\varphi(k,x)$, $g(k,x)$ and their $x$-derivatives $f'(k,x)$, $S'(k,x)$, $\varphi'(k,x)$, $g'(k,x)$ contain $k$ as $ik$, and hence they satisfy for $k \in \mathbf{C}$*

$$\psi(-k^*, x) = \psi(k, x)^*, \quad \psi'(-k^*, x) = \psi'(k, x)^*, \qquad x \in [0, \ell]. \tag{3.8}$$

(e) *The regular solution $\varphi(k,x)$ and its $x$-derivative $\varphi'(k,x)$ satisfy*

$$\varphi(-k, x) = \varphi(k, x), \quad \varphi'(-k, x) = \varphi'(k, x), \qquad k \in \mathbf{C}. \tag{3.9}$$

(f) *The quantity $\varphi(0,x)$ satisfies the zero-energy Schrödinger equation*

$$-\psi''(x) + q(x)\,\psi(x) = 0, \qquad x \in (0, \ell), \tag{3.10}$$

*and it is related to $r(x)$ as*

$$\varphi(0, x) = \frac{r(x)}{r_0}, \quad \varphi'(0, x) = \frac{r'(x)}{r_0}, \qquad x \in [0, \ell], \tag{3.11}$$

*and hence in particular we have*

$$\varphi(0, 0) = 1, \quad \varphi'(0, 0) = \frac{r'_0}{r_0}, \quad \varphi(0, \ell) = \frac{r_\ell}{r_0}, \quad \varphi'(0, \ell) = \frac{r'_\ell}{r_0}. \tag{3.12}$$

PROOF: We note that (a) directly follows from the properties of $r(x)$ listed in (a) and (c) of Definition 1.1. For each particular solution, the proof of (b) is obtained in the standard way by combining (3.2) and the relevant initial conditions into an integral equation and by showing that the resulting integral equation can be represented as a uniformly convergent infinite series. By using the fact that $q(x)$ is integrable, one can solve the corresponding integral equation iteratively to establish the uniform convergence. The proof of (c) is obtained as follows. In the proof of (b), one shows that each term in the infinite series is entire in $k$ and that the uniform convergence holds for $k \in \mathbf{C}$. The $x$-derivatives of the four particular solutions are also represented as uniformly convergent infinite series where each term is entire in $k$, and hence the Weierstrass theorem implies (c). The proof of (d) directly follows from the fact that $k$ appears as $ik$ both in the Schrödinger equation (3.2) and in each of the initial conditions (3.4)-(3.7). We obtain (e) by observing that $k$ appears as $k^2$ in (3.2) and (3.6). The proof of (f) is obtained as follows. By taking the $k$-derivative of (3.2) and inserting $k=0$ in the resulting equation, we get (3.10). Using the analyticity at $k=0$ of $\varphi(k,x)$ for each fixed $x \in [0,\ell]$, we observe that $\varphi(0,x)$ satisfies (3.10). Finally, we obtain (3.11) by using the fact that $\varphi(0,x)$ and $r(x)/r_0$ both satisfy the zero-energy Schrödinger equation (3.10) and the initial conditions (3.6) and that the corresponding initial-value problem has a unique solution. ∎

Let us define the Wronskian of two solutions $\psi(k,x)$ and $\phi(k,x)$ to (3.2) as

$$[\psi(k,x); \phi(k,x)] := \psi(k,x)\,\phi'(k,x) - \psi'(k,x)\,\phi(k,x). \tag{3.13}$$

It is already known and can also be directly verified that the Wronskian given in (3.13) is independent of $x$, and hence its value can be evaluated at $x=0$ or at $x=\ell$, yielding the same quantity. Consequently, using (3.5) and (3.6) in (3.13) we get

$$[\varphi(k,x); S(k,x)] = 1, \tag{3.14}$$

which indicates that $\varphi(k,x)$ and $S(k,x)$ are linearly independent for $x \in [0,\ell]$ for each fixed $k \in \mathbf{C}$.

Let us use $G(k)$ to denote the Wronskian of the solutions $\varphi(k,x)$ and $g(k,x)$ appearing in (3.6) and (3.7), respectively, i.e.

$$G(k) := [\varphi(k,x); g(k,x)]. \tag{3.15}$$

In the following proposition we list some useful properties of $G(k)$.

**Proposition 3.2** *Assume that the vocal-tract radius $r(x)$ belongs to the class $\mathcal{A}$ specified in Definition 1.1. Then, the quantity $G(k)$ defined in (3.15) has the following properties:*

(a) *It is entire and satisfies*

$$G(-k^*) = G(k)^*, \qquad k \in \mathbf{C}. \tag{3.16}$$

*Hence, the zeros of $G(k)$ are located symmetrically with respect to the imaginary axis in $\mathbf{C}$.*

(b) *It has a simple zero at $k = 0$ and we have*

$$G(0) = 0, \quad \dot{G}(0) = -i\frac{r_\ell}{r_0}. \tag{3.17}$$

*where an overdot denotes the $k$-derivative.*

(c) *The quantity $G(k)$ is nonzero when $k \in \mathbf{R} \setminus \{0\}$.*

(d) *The quantity $G(k)$ is nonzero when $k$ is on the imaginary axis in $\mathbf{C}$, except at $k = 0$ where it has a simple zero. In fact, $G(i\kappa) > 0$ for $\kappa \in \mathbf{R}^+$ and $G(i\kappa) < 0$ for $\kappa \in \mathbf{R}^-$.*

(e) *The quantity $G(k)$ and the quantity $z(k)$ appearing in (1.14) cannot simultaneously vanish at any point in $\mathbf{C} \setminus \{0\}$.*

(f) *If $G(k_0) = 0$ for some $k_0 \in \mathbf{C} \setminus \{0\}$, then $G(-k_0) = 0$ if and only if $J_1(2k_0 r_l) = k_0 r_l$.*

PROOF: We observe that (a) directly follows by using (c) and (d) of Proposition 3.1 in (3.15). From (3.16) we conclude that the zeros of $G(k)$ appear symmetrically with respect to the imaginary axis in $\mathbf{C}$. Thus, the proof of (a) is complete. The proof of (b) is obtained as follows. Evaluating the right-hand side of (3.15) at $x = \ell$ and using (3.7) we obtain

$$G(k) = \varphi(k, \ell) \left[ \frac{r'_\ell}{r_\ell} z(k) - ik \right] - \varphi'(k, \ell) z(k). \tag{3.18}$$

From (3.9), because $\varphi(k, \ell)$ and $\varphi'(k, \ell)$ are entire, it follows that

$$\varphi(k, \ell) = \varphi(0, \ell) + O(k^2), \quad \varphi'(k, \ell) = \varphi'(0, \ell) + O(k^2), \qquad k \to 0 \text{ in } \mathbf{C},$$

which, with the help of (3.12), can be written as

$$\varphi(k, \ell) = \frac{r_\ell}{r_0} + O(k^2), \quad \varphi'(k, \ell) = \frac{r'_\ell}{r_0} + O(k^2), \qquad k \to 0 \text{ in } \mathbf{C}. \tag{3.19}$$

From (2.2), (3.7), and the fact that $z(k)$ is entire, we get

$$g(k,\ell) = \frac{8ikr_\ell}{3\pi} + O(k^2), \qquad k \to 0 \text{ in } \mathbf{C}, \tag{3.20}$$

$$g'(k,\ell) = \left(\frac{8ir'_\ell}{3\pi} - i\right)k + O(k^2), \qquad k \to 0 \text{ in } \mathbf{C}. \tag{3.21}$$

Using (3.19)-(3.21) in (3.18) we obtain the expansion

$$G(k) = -\frac{r_\ell}{r_0} ik + O(k^2), \qquad k \to 0 \text{ in } \mathbf{C}, \tag{3.22}$$

which proves (b) by also establishing (3.17). The proof of (c) can be obtained as follows. From the properties [1] of the Struve function $\mathbf{H}_1(w)$, we already know that $\mathbf{H}_1(w) > 0$ when $w > 0$. Then, from (3.8) and (3.9) it follows that $\varphi(k,\ell)$ and $\varphi'(k,\ell)$ are real valued when $k \in \mathbf{R}$. Hence, using (3.18) we can separate the real and imaginary parts of $G(k)$ as

$$\mathrm{Re}[G(k)] = \left[\frac{r'_\ell}{r_\ell}\varphi(k,\ell) - \varphi'(k,\ell)\right]\mathrm{Re}[z(k)], \qquad k \in \mathbf{R}, \tag{3.23}$$

$$\mathrm{Im}[G(k)] = \left[\frac{r'_\ell}{r_\ell}\varphi(k,\ell) - \varphi'(k,\ell)\right]\mathrm{Im}[z(k)] - k\,\varphi(k,\ell), \qquad k \in \mathbf{R}. \tag{3.24}$$

From (d) and (e) of Proposition 2.1, we know that

$$\mathrm{Re}[z(k)] \neq 0, \quad \mathrm{Im}[z(k)] \neq 0, \qquad k \in \mathbf{R} \setminus \{0\}.$$

Thus, from (3.23) and (3.24) we conclude that $G(k)$ would vanish at a real nonzero value $k_0$ if and only if we had

$$\varphi(k_0,\ell) = 0, \quad \varphi'(k_0,\ell) = 0. \tag{3.25}$$

Since $\varphi(k_0,x)$ satisfies (3.2) with $k = k_0$, we would conclude from (3.25) that $\varphi(k_0,x) \equiv 0$, which contradicts the first equality in (3.6). Thus, we must have $G(k_0) \neq 0$, which completes the proof of (c). Let us now prove (d). Using (3.2) and (3.3) we get

$$\frac{d}{dx}\left[\varphi(k,x)\left(\varphi'(k,x) - \frac{r'(x)}{r(x)}\varphi(k,x)\right)\right] = -k^2\,\varphi(k,x)^2 + \left(\varphi'(k,x) - \frac{r'(x)}{r(x)}\varphi(k,x)\right)^2,$$

which, after using (3.6), yields

$$\int_0^\ell dx \left[ -k^2 \varphi(k,x)^2 + \left( \varphi'(k,x) - \frac{r'(x)}{r(x)} \varphi(k,x) \right)^2 \right]$$
$$= \varphi(k,\ell) \left( \varphi'(k,\ell) - \frac{r'_\ell}{r_\ell} \varphi(k,\ell) \right). \tag{3.26}$$

From (3.18) we have

$$\varphi'(k,\ell) - \frac{r'_\ell}{r_\ell} \varphi(k,\ell) = \frac{-ik}{z(k)} \varphi(k,\ell) - \frac{G(k)}{z(k)}. \tag{3.27}$$

Using (3.27) on the right-hand side of (3.26) we obtain

$$\int_0^\ell dx \left[ -k^2 \varphi(k,x)^2 + \left( \varphi'(k,x) - \frac{r'(x)}{r(x)} \varphi(k,x) \right)^2 \right]$$
$$= \frac{-ik}{z(k)} \varphi(k,\ell)^2 - \frac{G(k)}{z(k)} \varphi(k,\ell). \tag{3.28}$$

If we had $G(i\kappa_0) = 0$ for some real nonzero $\kappa_0$, then (3.28) would yield

$$\int_0^\ell dx \left[ \kappa_0^2 \varphi(i\kappa_0,x)^2 + \left( \varphi'(i\kappa_0,x) - \frac{r'(x)}{r(x)} \varphi(i\kappa_0,x) \right)^2 \right] = \frac{\kappa_0}{z(i\kappa_0)} \varphi(i\kappa_0,\ell)^2. \tag{3.29}$$

From Proposition 3.1(d) we know that $\varphi(i\kappa_0,x)$ and $\varphi'(i\kappa_0,x)$ are real valued, and hence the left-hand side of (3.29) is positive unless $\varphi(i\kappa_0,x) \equiv 0$, in which case that left-hand side is zero. From Proposition 2.1(f) we know that $\kappa_0/z(i\kappa_0)$ is negative. Hence, (3.29) could hold only when $\varphi(i\kappa_0,x) \equiv 0$, but that contradicts the first equality in (3.6). Thus, we must have $G(i\kappa_0) \neq 0$, which completes the proof of (d). Let us now prove (e). If we had $G(k_0) = z(k_0) = 0$ for some nonzero $k_0$ in $\mathbf{C}$, then we would see from (3.18) that $\varphi(k_0,\ell) = 0$. Thus, $\varphi(k_0,x)$ would be an eigenfunction with the eigenvalue $k_0^2$ for the operator associated with (3.2) on $x \in (0,\ell)$ with the real-valued potential $q(x)$ and the selfadjoint boundary conditions

$$\psi'(0) - \frac{r'_0}{r_0} \psi(0) = 0, \quad \psi(\ell) = 0. \tag{3.30}$$

The corresponding operator is selfadjoint and hence the eigenvalue $k_0^2$ would have to be real. This would require that $k_0$ would be either a real nonzero number or a purely imaginary nonzero number. However, from (c) and (d) we know that we must then have $G(k_0) \neq 0$. Thus, the proof of (e) is complete. Let us finally proceed with the proof of (f). Using (3.9) in (3.15) we see that

$$G(-k) = [\varphi(k,x); g(-k,x)], \qquad k \in \mathbf{C}. \tag{3.31}$$

From (3.15) we know that $G(k_0) = 0$ for some nonzero $k_0$ in $\mathbf{C}$ if and only if $\varphi(k_0, x)$ and $g(k_0, x)$ are linearly dependent. Similarly, from (3.31) we see that $G(-k_0) = 0$ if and only if $\varphi(k_0, x)$ and $g(-k_0, x)$ are linearly dependent. Thus, we have

$$G(k_0) = G(-k_0) = 0, \tag{3.32}$$

if and only if $\varphi(k_0, x)$, $g(k_0, x)$, and $g(-k_0, x)$ are pairwise linearly dependent. Since both $g(k_0, x)$ and $g(-k_0, x)$ are solutions to (3.2), their Wronskian can be evaluated at $x = \ell$. Hence, from (3.13) we get

$$[g(k_0, x); g(-k_0, x)] = g(k_0, \ell)\, g'(-k_0, \ell) - g(k_0, \ell)\, g'(-k_0, \ell),$$

which, with the help of (3.7), can be written as

$$[g(k_0, x); g(-k_0, x)] = ik_0 \left[z(k_0) + z(-k_0)\right].$$

Thus, the linear dependence of $g(k_0, x)$ and $g(-k_0, x)$ is equivalent to

$$z(k_0) + z(-k_0) = 0. \tag{3.33}$$

From (1.15) and (2.6) we see that (3.33) is equivalent to

$$1 - \frac{J_1(2k_0 r_l)}{k_0 r_l} = 0. \tag{3.34}$$

Thus, (3.32) yields (3.34). On the other hand, (3.34) implies that $g(k_0, x)$ and $g(-k_0, x)$ are linearly dependent and $G(k_0) = 0$ implies that $\varphi(k_0, x)$ and $g(k_0, x)$ are linearly dependent. Thus, (3.34) and $G(k_0) = 0$ together yield $G(-k_0) = 0$, which completes the proof of (f). ∎

From Proposition 3.1(c) we know that the regular solution $\varphi(k,x)$ and its derivative $\varphi'(k,x)$ are entire in $k$ for each fixed $x \in [0,\ell]$. In the next proposition we establish their large-$k$ asymptotics in $\mathbf{C}$.

**Proposition 3.3** *Assume that the vocal-tract radius $r(x)$ belongs to the class $\mathcal{A}$ specified in Definition 1.1. Let $\varphi(k,x)$ be the regular solution to the Schrödinger equation (3.2) satisfying the initial conditions given in (3.6). We then have the following:*

(a) *The quantity $e^{\pm ikx}\varphi(k,x)$ satisfies*

$$e^{\pm ikx}\varphi(k,x) = \frac{1}{2}\left[1 + e^{\pm 2ikx}\right] \mp \frac{r_0'}{2ikr_0}\left[1 - e^{\pm 2ikx}\right]$$
$$\mp \frac{1}{4ik}\int_0^x dy\left[1 - e^{\pm 2ikx} + e^{\pm 2iky} - e^{\pm 2ik(x-y)}\right]q(y) \qquad (3.35)$$
$$+ O\left(\frac{1}{k^2}\right), \qquad k \to \infty \text{ in } \overline{\mathbf{C}^{\pm}}.$$

(b) *The quantity $e^{\pm ikx}\varphi'(k,x)$ satisfies*

$$e^{\pm ikx}\varphi'(k,x) = \pm\frac{k}{2i}\left[1 - e^{\pm 2ikx}\right] + \frac{r_0'}{2r_0}\left[1 + e^{\pm 2ikx}\right]$$
$$+ \frac{1}{4}\int_0^x dy\left[1 + e^{\pm 2ikx} + e^{\pm 2iky} + e^{\pm 2ik(x-y)}\right]q(y) \qquad (3.36)$$
$$+ O\left(\frac{1}{k}\right), \qquad k \to \infty \text{ in } \overline{\mathbf{C}^{\pm}}.$$

(c) *Consequently, as $k \to \infty$ in $\overline{\mathbf{C}^+}$ we have*

$$e^{ik\ell}\varphi(k,\ell) = \frac{1}{2}\left(1 + e^{2ik\ell}\right) + O\left(\frac{1}{k}\right), \qquad k \to \infty \text{ in } \overline{\mathbf{C}^+}, \qquad (3.37)$$

$$e^{ik\ell}\varphi'(k,\ell) = \frac{k}{2i}\left(1 - e^{2ik\ell}\right) + O(1), \qquad k \to \infty \text{ in } \overline{\mathbf{C}^+}, \qquad (3.38)$$

*and as $k \to \infty$ in $\overline{\mathbf{C}^-}$ we have*

$$e^{-ik\ell}\varphi(k,\ell) = \frac{1}{2}\left(1 + e^{-2ik\ell}\right) + O\left(\frac{1}{k}\right), \qquad k \to \infty \text{ in } \overline{\mathbf{C}^-}, \qquad (3.39)$$

$$e^{-ik\ell}\varphi'(k,\ell) = -\frac{k}{2i}\left(1 - e^{-2ik\ell}\right) + O(1), \qquad k \to \infty \text{ in } \overline{\mathbf{C}^-}. \qquad (3.40)$$

PROOF: With the help of (3.6), since $\varphi(k,x)$ satisfies (3.2), we get the integral relations

$$\varphi(k,x) = \cos kx + \frac{r_0'}{r_0}\frac{\sin kx}{k} + \frac{1}{k}\int_0^x dy\,[\sin k(x-y)]\,q(y)\,\varphi(k,y), \qquad (3.41)$$

$$\varphi'(k,x) = -k\sin kx + \frac{r_0'}{r_0}\cos kx + \int_0^x dy\,[\cos k(x-y)]\,q(y)\,\varphi(k,y). \qquad (3.42)$$

Multiplying both sides of (3.41) and (3.42) with $e^{\pm ikx}$, we can write the resulting integral relations in terms of $e^{\pm ikx}\varphi(k,x)$ and $e^{\pm ikx}\varphi'(k,x)$. Using iteration on the resulting equations we obtain the large-$k$ asymptotics given in (3.35) and (3.36). We remark that all the exponential terms appearing on the right-hand sides of (3.35) and (3.36) are bounded in the appropriate upper or lower half of the complex plane. Using the leading asymptotics in (3.35) and (3.36) as $k \to \infty$ in $\overline{\mathbf{C}^+}$ we get (3.37) and (3.38). Similarly, by using the leading asymptotics in (3.35) and (3.36) as $k \to \infty$ in $\overline{\mathbf{C}^-}$ we get (3.39) and (3.40). ∎

With the help of the large $k$-asymptotics for $z(k)$ established in Proposition 2.2 and large $k$-asymptotics for $\varphi(k,\ell)$ and $\varphi'(k,\ell)$ established in Proposition 3.3, in the next theorem we establish the large-$k$ asymptotics of $G(k)$ appearing in (3.18).

**Proposition 3.4** *Assume that the vocal-tract radius $r(x)$ belongs to the class $\mathcal{A}$ specified in Definition 1.1. Let $G(k)$ be the quantity appearing in (3.15) and (3.18). We then have the following:*

(a) *The large-$k$ behavior of $G(k)$ in $\mathbf{C}^+$ is given by*

$$\begin{aligned} e^{ik(2r_\ell+\ell)}\,G(k) &= \frac{1+i}{2r_\ell\sqrt{2\pi k r_\ell}}\left[1+O\left(\frac{1}{k}\right)\right] + e^{2ikr_\ell}\,O(k) \\ &\quad + e^{2ik\ell}\,O\left(\frac{1}{|k|^{1/2}}\right), \qquad k \to \infty \text{ in } \mathbf{C}^+. \end{aligned} \qquad (3.43)$$

(b) *The large-$k$ behavior of $G(k)$ in $\mathbf{C}^-$ and in $\mathbf{R}$ is determined by*

$$e^{-ik\ell}\,G(k) = -ik + O(1) + e^{-2ik\ell}\,O(k), \qquad k \to \infty \text{ in } \overline{\mathbf{C}^-}. \qquad (3.44)$$

(c) *The entire function $G(k)$ is of order one, i.e. it is of exponential type.*

(d) *The large-k behavior of $G(k)$ in $\mathbf{R}$ is given by*

$$G(k) = \left[-ik - \gamma + \frac{r'_\ell}{r_\ell}\right] \cos k\ell + \left[k - i\gamma + \frac{2i}{\pi r_\ell}\right] \sin kl + o(1), \qquad k \to \pm\infty, \quad (3.45)$$

*where we have defined the constant $\gamma$ as*

$$\gamma := \frac{r'_0}{r_0} + \frac{1}{2}\int_0^\ell dy\, q(y). \qquad (3.46)$$

(e) *The large-k behaviors of $|G(k)|$ and the logarithm of $|G(k)|$ in $\mathbf{R}$ are given by*

$$\frac{|G(k)|}{|k|} = 1 + O\left(\frac{1}{k}\right), \qquad k \to \pm\infty, \qquad (3.47)$$

$$\ln|G(k)| = \ln|k| + o(1), \qquad k \to \pm\infty. \qquad (3.48)$$

(f) *The quantity $(\ln|G(k)|)/(1+k^2)$ is integrable at $k = \pm\infty$, and we have*

$$\int_{-\infty}^\infty dk\, \frac{\max\{0, |\ln|G(k)||\}}{1+k^2} < +\infty. \qquad (3.49)$$

(g) *The entire function $G(k)$ belongs to the Cartwright class, as defined on p. 97 of [14].*

(h) *There are at most a finite number of zeros of $G(k)$ in $\mathbf{C}^-$.*

(i) *The quantity $G(k)$ satisfies*

$$\lim_{\kappa \to +\infty} \left[\frac{\ln|G(i\kappa)|}{\kappa}\right] = 2r_\ell + \ell, \qquad (3.50)$$

$$\lim_{\kappa \to +\infty} \left[\frac{\ln|G(-i\kappa)|}{\kappa}\right] = \ell. \qquad (3.51)$$

PROOF: We obtain (3.44) by using (2.12), (3.37), and (3.38) in (3.18) and by keeping the leading terms as $k \to \infty$ in $\mathbf{C}^+$. Similarly, we obtain (3.44) by using (2.13), (3.39), and (3.40) in (3.18) and by keeping the leading terms as $k \to \infty$ in $\overline{\mathbf{C}^-}$. Let us now prove (c). The exponential terms $e^{2ikr_\ell}$ and $e^{2ik\ell}$ appearing on the right-hand side of (3.43) are bounded in $k \in \mathbf{C}^+$. Similarly, the exponential term $e^{-2ik}$ appearing on the right-hand side of (3.44) is bounded in $k \in \overline{\mathbf{C}^-}$. From the large-k asymptotics in $\mathbf{C}$ given in (3.43) and (3.44), we observe that the entire function $G(k)$ has order one and hence it is of exponential type. Thus, the proof of (c) is complete. We obtain (d) as follows. We multiply (3.35) by $e^{\mp ikx}$ on both sides and evaluate the resulting expansion at $x = \ell$ and apply the Riemann-Lebesgue lemma and obtain

$$\varphi(k, \ell) = \cos k\ell + \frac{\gamma \sin k\ell}{k} + o\left(\frac{1}{k}\right), \qquad k \to \pm\infty. \tag{3.52}$$

By a similar procedure from (3.36) we get

$$\varphi'(k, \ell) = -k \sin k\ell + \gamma \cos k\ell + o(1), \qquad k \to \pm\infty. \tag{3.53}$$

Finally, using (2.13), (3.52), and (3.53) in (3.18) we obtain (3.45). Thus, the proof of (d) is complete. Next, from the behavior in (3.44) as $k \to \pm\infty$ in $\mathbf{R}$, we obtain (3.47) and (3.48), completing the proof of (e). Let us now turn the proof of (f). Because $G(k)$ is entire, $|G(k)|$ is continuous in $k \in \mathbf{R}$. From (3.48) we observe that $(\ln|G(k)|)/(1+k^2)$ is integrable at $k = \pm\infty$ and hence (3.49) holds, which completes the proof of (f). Being entire, exponential type, and satisfying (3.49), by the definition of the Cartwright class [14] it follows that $G(k)$ satisfies the property stated in (g). Let us now prove (h). From (a), (b), and (c) of Proposition 3.2 we conclude that $i\,e^{-ik\ell}G(k)/k$ is an entire function and does not vanish for $k \in \mathbf{R}$; furthermore, from (3.44) we see that it has the behavior $1 + O(1/k)$ as $k \to \infty$ in $\overline{\mathbf{C}^-}$. Thus, it is analytic in $\mathbf{C}^-$, continuous in $\overline{\mathbf{C}^-}$, and does not vanish on the boundary of $\mathbf{C}^-$. Hence, it cannot have infinitely many zeros in $\mathbf{C}^-$. Consequently, $G(k)$ cannot have infinitely many zeros in $\mathbf{C}^-$, which completes the proof of (h). In fact, from (a), (b), and (d) of Proposition 3.2 we know that $G(k)$ has a simple zero at $k = 0$, has no other zeros on the imaginary axis, and all its nonzero zeros are located symmetrically with respect to the imaginary axis. Let us finally prove (i). Letting $k = i\kappa$, from (3.43) we get

$$-\kappa(2r_\ell + \ell) + \ln|G(i\kappa)| = O(\ln|k|), \qquad \kappa \to +\infty,$$

which yields (3.50). Similarly, letting $k = -i\kappa$, from (3.44) we obtain

$$e^{-\kappa\ell} G(-i\kappa) = -\kappa\left[1 + O\left(\frac{1}{\kappa}\right)\right], \qquad \kappa \to +\infty,$$

which yields (3.51). Hence, the proof of (i) is complete. We remark that, as indicated in Proposition 3.2(d), we can also use $\ln(G(i\kappa))$ in (3.50) instead of $\ln|G(i\kappa)|$ and we can use $\ln(-G(-i\kappa))$ in (3.51) instead of $\ln|G(-i\kappa)|$. ∎

## 4. FURTHER PROPERTIES OF $G(k)$

In this section we present some further relevant properties of the key quantity $G(k)$ appearing in (3.15) and (3.18).

Note that the real and imaginary axes divide $\mathbf{C}$ into four quadrants. Let us use $\{k_j^-\}_{j=1}^{N^-}$ to denote the set of zeros of $G(k)$ in the fourth quadrant, where we allow the multiplicities of the zeros in our count and we order the zeros so that $|k_j^-| \leq |k_{j+1}^-|$. Thus, we have

$$\operatorname{Re}[k_j^-] > 0, \quad \operatorname{Im}[k_j^-] < 0, \qquad 1 \leq j \leq N^-. \tag{4.1}$$

By Proposition 3.4(h) we know that $N^-$ is finite. Furthermore, by Proposition 3.2(a) we conclude that the set of zeros of $G(k)$ in the third quadrant is given by $\{-(k_j^-)^*\}_{j=1}^{N^-}$. Similarly, let us use $\{k_j^+\}_{j=1}^{N^+}$ to denote the set of zeros of $G(k)$ in the first quadrant, where we allow the multiplicities of the zeros in our count and we order the zeros so that $|k_j^+| \leq |k_{j+1}^+|$. Thus, we have

$$\operatorname{Re}[k_j^+] > 0, \quad \operatorname{Im}[k_j^+] > 0, \qquad 1 \leq j \leq N^+. \tag{4.2}$$

By Proposition 3.2(a) we conclude that the set of zeros of $G(k)$ in the second quadrant is given by $\{-(k_j^+)^*\}_{j=1}^{N^+}$. The next proposition shows that $N^+ = +\infty$ and also describes the density of the zeros in the set $\{k_j^+\}_{j=1}^{N^+}$.

**Proposition 4.1** *Assume that the vocal-tract radius $r(x)$ belongs to the class $\mathcal{A}$ specified in Definition 1.1. The quantity $G(k)$ appearing in (3.15) and (3.18) has the following properties:*

(a) *There are infinitely many zeros of $G(k)$ in $\mathbf{C}^+$, and hence $N^+ = +\infty$.*

(b) *The zeros of $G(k)$ in the first quadrant of $\mathbf{C}$ satisfy*

$$\sum_{j=1}^{N^+} \left| \text{Im} \left[ \frac{1}{k_j^+} \right] \right| < +\infty. \tag{4.3}$$

(c) *The density of the zeros of $G(k)$ in the first quadrant as $k \to \infty$ is given by $(r_\ell + \ell)/\pi$. In other words, the number of zeros $k_j^+$ of $G(k)$ satisfying $\rho \leq |k_j^+| \leq \rho + 1$ in the limit $\rho \to +\infty$ is equal to $(r_\ell + \ell)/\pi$.*

PROOF: For each fixed positive $\rho$, let us use $n_+(\rho)$ to denote the number of zeros of $G(k)$ with positive real parts and lying in the disk $|k| \leq \rho$ in $\mathbf{C}$. Thus, $n_+(\rho)$ is the number of zeros of $G(k)$ satisfying

$$0 < |k_j^-| \leq \rho, \quad 0 < |k_j^+| \leq \rho.$$

From Proposition 3.4(g), we know that $G(k)$ belongs to the Cartwright class. Consequently, the Cartwright-Levinson theorem, see e.g. p. 127 of [14], holds and we have

$$\lim_{\rho \to +\infty} \frac{n_+(\rho)}{\rho} = \frac{1}{2\pi} \left[ \lim_{\kappa \to +\infty} \frac{\ln |G(i\kappa)|}{\kappa} + \lim_{\kappa \to +\infty} \frac{\ln |G(-i\kappa)|}{\kappa} \right]. \tag{4.4}$$

Using (3.50) and (3.51) on the right-hand side of (4.4) we get

$$\lim_{\rho \to +\infty} \frac{n_+(\rho)}{\rho} = \frac{r_\ell + \ell}{\pi},$$

which yields

$$n_+(\rho) = \frac{r_\ell + \ell}{\pi} \rho \left[ 1 + o(1) \right], \quad \rho \to +\infty. \tag{4.5}$$

From (4.5) we get

$$N^- + N^+ = +\infty. \tag{4.6}$$

The finiteness of $N^-$ is known from Proposition 3.4(h), and hence from (4.6) we conclude that $N^+ = +\infty$. Since the zeros of $G(k)$ are located symmetrically with respect to the imaginary axis, as stated in Proposition 3.2(a), the number of zeros of $G(k)$ in $\mathbf{C}^+$, which is equal to $2N^+$, must be infinite. Thus, the proof of (a) is complete. Let us now prove (b). From the Cartwright-Levinson theorem we know that the finiteness in (4.3) holds when we include all nonzero zeros of $G(k)$ on the left-hand side of (4.3). In other words, the Cartwright-Levinson theorem yields

$$\sum_{j=1}^{N^+}\left|\operatorname{Im}\left[\frac{1}{k_j^+}\right]\right|+\sum_{j=1}^{N^+}\left|\operatorname{Im}\left[\frac{1}{(-(k_j^+)^*}\right]\right|+\sum_{j=1}^{N^-}\left|\operatorname{Im}\left[\frac{1}{k_j^-}\right]\right|+\sum_{j=1}^{N^-}\left|\operatorname{Im}\left[\frac{1}{-(k_j^-)^*}\right]\right|<+\infty,$$

which is equivalent to

$$2\sum_{j=1}^{N^+}\left|\operatorname{Im}\left[\frac{1}{k_j^+}\right]\right|+2\sum_{j=1}^{N^-}\left|\operatorname{Im}\left[\frac{1}{k_j^-}\right]\right|<+\infty,$$

which in turn implies (4.3). Let us finally prove (c). Using $\rho+1$ in (4.5) we obtain

$$n_+(\rho+1)=\frac{r_\ell+\ell}{\pi}(\rho+1)\left[1+o(1)\right],\qquad\rho\to+\infty. \tag{4.7}$$

Subtracting (4.7) from (4.5) we get

$$n_+(\rho+1)-n_+(\rho)=\frac{r_\ell+\ell}{\pi}+o(1),\qquad\rho\to+\infty,$$

which establishes (c). ∎

In the next proposition we discuss the construction of $G(k)$ for $k\in\mathbf{C}$ from its zeros.

**Proposition 4.2** *Assume that the vocal-tract radius $r(x)$ belongs to the class $\mathcal{A}$ specified in Definition 1.1. Let $G(k)$ be the corresponding quantity appearing in (3.15) and (3.18). Let us use $\{k_j^+\}_{j=1}^{N^+}$ and $\{k_j^-\}_{j=1}^{N^-}$ to denote the set of zeros of $G(k)$ in the first and fourth quadrants, respectively, where $N^-$ is a nonnegative integer and $N^+=+\infty$. If we know the values of $\ell$ as well as the sets $\{k_j^+\}_{j=1}^{N^+}$ and $\{k_j^-\}_{j=1}^{N^-}$, then we can uniquely and explicitly construct $G(k)$ for $k\in\mathbf{C}$.*

PROOF: By Proposition 3.2(a) we know that $G(k)$ is entire and by Proposition 3.4(c) we know that $G(k)$ is of exponential type. Hence, the Hadamard factorization of $G(k)$ can be written as

$$G(k)=\frac{r_\ell}{r_0}\,A\,e^{iCk}\,E(k), \tag{4.8}$$

where $A$ and $C$ are some constants and the quantity $E(k)$ is uniquely determined by the zeros of $G(k)$ as

$$E(k):=-ik\,E^-(k)\,E^+(k), \tag{4.9}$$

$$E^-(k) := \prod_{j=1}^{N^-} \left(1 - \frac{k}{k_j^-}\right) e^{k/k_j^-} \left(1 + \frac{k}{(k_j^-)^*}\right) e^{-k/(k_j^-)^*}, \tag{4.10}$$

$$E^+(k) := \prod_{j=1}^{N^+} \left(1 - \frac{k}{k_j^+}\right) e^{k/k_j^+} \left(1 + \frac{k}{(k_j^+)^*}\right) e^{-k/(k_j^+)^*}, \tag{4.11}$$

where we have used the fact that $G(k)$ has no zeros on the real and imaginary axes except for a simple zero at $k = 0$ and that the zeros of $G(k)$ are symmetrically located with respect to the imaginary axis, as stated in Proposition 3.2. We remark that $E(k)$ specified in (4.9)-(4.11) satisfies

$$E(-k^*) = E(k)^*, \qquad k \in \mathbf{C}, \tag{4.12}$$

and hence $E(k)$ is real valued when $k$ is on the imaginary axis. In fact, the choice of the factor $-i$ in (4.9) is for the convenience that the sign of $E(i\kappa)$ coincides with the sign of $\kappa$ when $\kappa \in \mathbf{R}$. This follows from the fact that $E^-(i\kappa)$ and $E^+(i\kappa)$ are both positive for $\kappa \in \mathbf{R}$ because from (4.9)-(4.11) we get

$$E(i\kappa) = \kappa\, E^-(i\kappa)\, E^+(i\kappa), \tag{4.13}$$

$$E^-(i\kappa) = \prod_{j=1}^{N^-} \left(1 + \frac{\kappa^2}{|k_j^-|^2}\right) \exp\left(\frac{2\kappa \operatorname{Im}[k_j^-]}{|k_j^-|^2}\right),$$

$$E^+(i\kappa) = \prod_{j=1}^{N^+} \left(1 + \frac{\kappa^2}{|k_j^+|^2}\right) \exp\left(\frac{2\kappa \operatorname{Im}[k_j^+]}{|k_j^+|^2}\right).$$

We will show that the constants $A$ and $C$ are determined by $\ell$ and $E(k)$. From (3.22) we know that

$$\lim_{k \to 0} \frac{G(k)}{k} = -i\frac{r_\ell}{r_0}. \tag{4.14}$$

Using (4.8)-(4.11) on the left-hand side of (4.14) we see that $A = 1$, and hence (4.8) is equivalent to

$$G(k) = \frac{r_\ell}{r_0} e^{iCk} E(k). \tag{4.15}$$

We remark that $G(k)$ satisfies (3.16) and $E(k)$ satisfies (4.12), and hence we conclude from (4.15) that the value of $C$ is real. Let us now show that $C$ is uniquely determined by $\ell$ and $E(k)$. Letting $k = i\kappa$ for $\kappa \in \mathbf{R}$ in (4.15) we get

$$G(i\kappa) = \frac{r_\ell}{r_0} e^{-C\kappa} E(i\kappa), \qquad \kappa \in \mathbf{R}. \tag{4.16}$$

From (4.16) we have

$$|G(i\kappa)| = \frac{r_\ell}{r_0} e^{-C\kappa} |E(i\kappa)|, \qquad \kappa \in \mathbf{R}, \tag{4.17}$$

or equivalently

$$\ln |G(i\kappa)| = \ln\left(\frac{r_\ell}{r_0}\right) - C\kappa + \ln |E(i\kappa)|, \qquad \kappa \in \mathbf{R}, \tag{4.18}$$

where we recall that $r_0$ and $r_\ell$ are both positive, as stated in Definition 1.1(a). From (4.18) we get

$$C = \frac{1}{\kappa} \ln\left(\frac{r_\ell}{r_0}\right) - \frac{\ln |G(i\kappa)|}{\kappa} + \frac{\ln |E(i\kappa)|}{\kappa}, \qquad \kappa \in \mathbf{R} \setminus \{0\}. \tag{4.19}$$

Letting $\kappa \to -\infty$ in (4.19) and using (3.51) we obtain

$$C = \ell + \lim_{\kappa \to -\infty} \left[\frac{\ln |E(i\kappa)|}{\kappa}\right]. \tag{4.20}$$

Thus, (4.20) shows that $C$ is uniquely determined by the value of $\ell$ and the zeros of $G(k)$. Using the right-hand side of (4.20) in (4.15) we conclude that $G(k)$ is determined by the values of $\ell$ and $r_\ell/r_0$ as well as the zeros of $G(k)$. The following argument shows that both $r_0$ and $r_\ell$ are determined by $E(k)$. Letting $\kappa \to +\infty$ in (4.19) and using (3.50) we obtain

$$C = -2r_\ell - \ell + \lim_{\kappa \to +\infty} \left[\frac{\ln |E(i\kappa)|}{\kappa}\right]. \tag{4.21}$$

Hence, equating the right-hand sides of (4.20) and (4.21) we obtain

$$r_\ell = -\ell + \frac{1}{2} \lim_{\kappa \to +\infty} \left[\frac{\ln |E(-i\kappa)| + \ln |E(i\kappa)|}{\kappa}\right]. \tag{4.22}$$

Thus, $r_\ell$ is determined by the knowledge of $\ell$ and $E(k)$. Let us also show that $r_0/r_\ell$ is also determined by the large $k$-asymptotics of $E(k)$ in $\mathbf{R}$. From (4.15), since $r_0$ and $r_\ell$ are positive and $C$ is real, we get

$$\frac{r_\ell}{r_0} = \frac{|G(k)|}{|E(k)|}, \qquad k \in \mathbf{R}.$$

Thus, we get

$$\frac{r_\ell}{r_0} = \frac{|G(k)|}{|k|} \frac{|k|}{|E(k)|}, \qquad k \in \mathbf{R}. \tag{4.23}$$

Letting $k \to \pm\infty$ in (4.23) and using (3.47) we obtain

$$\frac{r_\ell}{r_0} = \lim_{k \to \pm\infty} \left[ \frac{|k|}{|E(k)|} \right]. \tag{4.24}$$

Since we already know $r_\ell$, from (4.24) we see that $r_0$ is also determined. Then, from (4.15), (4.20), and (4.22) we conclude that $G(k)$ is uniquely and explicitly determined by the knowledge of $\ell$ and the zeros of $G(k)$. ∎

In the next proposition we discuss the explicit construction of $G(k)$ for $k \in \overline{\mathbf{C}^-}$ when we know $|G(k)|$ for $k \in \mathbf{R}^+$ and the zeros of $G(k)$ in $\mathbf{C}^-$.

**Proposition 4.3** *Assume that the vocal-tract radius $r(x)$ belongs to the class $\mathcal{A}$ specified in Definition 1.1. Let $G(k)$ be the corresponding quantity appearing in (3.15) and (3.18). Let us use $\{k_j^+\}_{j=1}^{N^+}$ and $\{k_j^-\}_{j=1}^{N^-}$ to denote the set of zeros of $G(k)$ in the first and fourth quadrants, respectively, where $N^-$ is a nonnegative integer and $N^+ = +\infty$.*

(a) *The knowledge of $G(k)$ for $k \in \mathbf{R}^+$ is sufficient to determine $G(k)$ for $k \in \mathbf{C}$ and in particular its zeros in $\mathbf{C}$.*

(b) *The quantity $G(k)$ for $k \in \overline{\mathbf{C}^-}$ is explicitly determined from the knowledge of $|G(k)|$ for $k \in \mathbf{R}^+$ and the set $\{k_j^-\}_{j=1}^{N^-}$ of zeros of $G(k)$ in the fourth quadrant.*

PROOF: As stated in Proposition 3.2(a), the quantity $G(k)$ is entire and hence by analytic extension from $k \in \mathbf{R}^-$ to $\mathbf{C}$, we see that $G(k)$ is uniquely determined for $k \in \mathbf{C}$. Thus, (a) is valid. Let us now prove (b). Let us introduce the quantity $H(k)$ in terms of $G(k)$ as

$$H(k) := e^{ik\ell} \frac{G(-k)}{ik}, \qquad k \in \mathbf{C}. \tag{4.25}$$

By Proposition 3.2(b) we know that $G(k)$ has a simple zero at $k = 0$ and hence $H(k)$ is entire and that $H(k)$ does not vanish at $k = 0$. From (3.16) it follows that

$$H(-k) = H(k)^*, \qquad k \in \mathbf{R},$$

and hence $|H(-k)| = |H(k)|$ for $k \in \mathbf{R}$, indicating that we know $|H(k)|$ for $k \in \mathbf{R}$ when $|G(k)|$ is known for $k \in \mathbf{R}^+$. From Proposition 3.2 and Proposition 3.4(h), by using the fact that the set of zeros of $G(k)$ in $k \in \overline{\mathbf{C}^-} \setminus \{0\}$ is given by $\{k_j^-, -(k_j^-)^*\}_{j=1}^{N_-}$, we conclude that the set of zeros of $H(k)$ in $k \in \overline{\mathbf{C}^+}$ is given by $\{-k_j^-, (k_j^-)^*\}_{j=1}^{N_-}$. Let us define

$$R(k) := \prod_{j=1}^{N^-} \frac{(k - k_j^-)(k + (k_j^-)^*)}{(k + k_j^-)(k - (k_j^-)^*)}. \tag{4.26}$$

We observe that $|R(k)| = 1$ for $k \in \mathbf{R}$ and we have $R(k)^{-1} = R(-k)$ for $k \in \mathbf{C}$. Note that $R(k) H(k)$ is analytic in $k \in \mathbf{C}^+$, continuous in $k \in \overline{\mathbf{C}^+}$, has no zeros in $k \in \overline{\mathbf{C}^+}$, and we have $|R(k) H(k)| = |H(k)|$ for $k \in \mathbf{R}$. Furthermore, from (3.44), (4.25), and (4.26) we conclude that $R(k) H(k)$ has the behavior $1 + O(1/k)$ as $k \to \infty$ in $\overline{\mathbf{C}^+}$. Thus, we can construct $R(k) H(k)$ in $k \in \overline{\mathbf{C}^+}$ from its absolute value known in $k \in \mathbf{R}$. The construction is analogous to the construction [10] of the transmission coefficient in the one-dimensional Schrödinger equation and we have

$$R(k) H(k) = \exp\left(\frac{1}{\pi i} \int_{-\infty}^{\infty} dy \, \frac{\ln |H(y)|}{y - k - i0^+}\right), \qquad k \in \overline{\mathbf{C}^+}, \tag{4.27}$$

where the presence of $i0^+$ indicates that when $k \in \mathbf{R}$ the right-hand side of (4.27) must be evaluated in the limit where $k \in \mathbf{C}^+$ moves to $k \in \mathbf{R}$. With the help of (4.25) and (4.26) and using the properties of $H(k)$, from (4.27) we get

$$e^{ik\ell} G(-k) = ik \, R(-k) \exp\left(\frac{1}{\pi i} \int_{-\infty}^{\infty} dy \, \frac{\ln |G(-y)/y|}{y - k - i0^+}\right), \qquad k \in \overline{\mathbf{C}^+}. \tag{4.28}$$

Changing the dummy integration variable from $y$ to $-y$ in (4.28) we obtain the explicit expression for $G(k)$ in $k \in \overline{\mathbf{C}^-}$ given by

$$G(k) = -ik \, e^{ik\ell} R(k) \exp\left(\frac{1}{\pi i} \int_{-\infty}^{\infty} dy \, \frac{\ln |G(y)/y|}{y - k + i0^+}\right), \qquad k \in \overline{\mathbf{C}^-}. \tag{4.29}$$

Thus, the proof of (b) is complete. ∎

In the next proposition we show that the knowledge of the key quantity $G(k)$ for $k \in \mathbf{R}^+$ along with the constant $\ell$ is sufficient to determine the constants $r_0$, $r_0'$, $r_\ell$, $r_\ell'$, and the normalized impedance $z(k)$.

**Proposition 4.4** *Assume that the vocal-tract radius $r(x)$ belongs to the class $\mathcal{A}$ with $r_0$, $r_0'$, $r_\ell$, $r_\ell'$ being the constants specified in Definition 1.1. Let $G(k)$ be the corresponding quantity appearing in (3.15) and (3.18). Let $z(k)$ be the normalized impedance given in (1.15) and let $q(x)$ be the potential defined in (3.3). Then, using the knowledge of $\ell$ and $G(k)$ for $k \in \mathbf{R}^+$ as input, we can construct $r_0$, $r_0'$, $r_\ell$, $r_\ell'$, and $z(k)$.*

PROOF: From Proposition 4.3(a) we know that $G(k)$ for $k \in \mathbf{C}$ including all its zeros is uniquely determined from the knowledge of $G(k)$ for $k \in \mathbf{R}^+$. Thus, we know $E(k)$ defined in (4.9). Then, as seen from (4.22), we can construct $r_\ell$ by using $\ell$ and $E(k)$. From (4.20) we see that, by using $\ell$ and $E(k)$, we can construct the constant $C$ appearing and (4.15). Since $G(k)$ is known, from (4.17) and (4.20) we observe that $r_0$ is also determined. Next, we obtain the constant $\gamma$ defined in (3.46) and the constant $r_\ell'$ from the large-$k$ limit of $G(k)$ described in (3.45). This can be achieved by letting $k = 2n\pi/\ell$ with $n \to +\infty$, and hence with $\cos k\ell = 1$ and $\sin k\ell = 0$ in (3.46), we obtain $-\gamma + r_\ell'/r_\ell$. Then, by letting $k = 2n\pi/\ell + \pi/(2\ell)$ with $n \to +\infty$, and hence with $\cos k\ell = 0$ and $\sin k\ell = 1$ in (3.46), we obtain $-\gamma + 2\pi/r_\ell$. Since we have already determined $r_\ell$, we then obtain both $r_\ell'$ and $\gamma$. Next, having the value of $r_\ell$ and using (1.15), we obtain the normalized impedance $z(k)$. ∎

A key quantity used in the inverse scattering theory related to the Schrödinger equation on the half line is the Jost function. The Jost function is usually denoted by $F(k)$ and is defined as a Wronskian as

$$F(k) := -i[\varphi(k,x); f(k,x)], \tag{4.30}$$

where we recall that $\varphi(k,x)$ is the regular solution to (3.2) satisfying (3.6) and $f(k,x)$ is the Jost solution satisfying (3.7), with the understanding that the potential $q(x)$ appearing in (3.3) is viewed in $x \in \mathbf{R}^+$ with compact support in $[0, \ell]$, and this is achieved by using the mathematical convention

$$r(x) = r_\ell'(x - \ell) + r_\ell, \qquad x \geq \ell.$$

Since the Wronskian in (4.30) is independent of $x$, its value can be evaluated at $x = 0$,

yielding
$$F(k) = -i\left(f'(k,0) - \frac{r'_0}{r_0}f(k,0)\right), \tag{4.31}$$

or it can be evaluated at $x = \ell$, yielding
$$F(k) = -i\,e^{ik\ell}\left(ik\,\varphi(k,\ell) - \varphi'(k,\ell)\right). \tag{4.32}$$

The Jost function is used as input to the Gel'fand-Levitan integral equation to obtain the potential and the boundary condition, which in our case, correspond to $q(x)$ given in (3.3) and the value of the constant $r'_0/r_0$, respectively.

As seen from (4.30), the Jost function $F(k)$ is affected by the constants $r_0$ and $r'_0$ but not by $r_\ell$ or $r'_\ell$. In other words, the boundary condition (1.11) at $x = \ell$ and the boundary condition (1.14) at $x = \ell$ yield the same Jost function $F(k)$. Thus, all the properties of $F(k)$ presented in [5] still hold when we use (1.14) instead of (1.11).

In the next proposition we show that the Jost function $F(k)$ can be constructed from the key quantity $G(k)$.

**Proposition 4.5** *Assume that the vocal-tract radius $r(x)$ belongs to the class $\mathcal{A}$ specified in Definition 1.1. Let $G(k)$ be the corresponding quantity appearing in (3.15) and (3.18). Let $F(k)$ be the Jost function appearing in (4.30)-(4.32), and let $q(x)$ be the potential defined in (3.3). Then, using the knowledge of $\ell$ and $G(k)$ for $k \in \mathbf{R}^+$ as input, we can construct $F(k)$ and $q(x)$.*

PROOF: From Proposition 4.3(a) we know that the knowledge of $\ell$ and $G(k)$ for $k \in \mathbf{R}^+$ yields the knowledge of $G(k)$ for $k \in \mathbf{C}$. We also know from Proposition 4.4 that the constants $r_\ell$ and $r'_\ell$ as well as the quantity $z(k)$ are also constructed from $\ell$ and $G(k)$. We can then view (3.23) and (3.24) as a linear algebraic system for the unknowns $\varphi(k,\ell)$ and $\varphi'(k,\ell)$, and we obtain its unique solution as

$$\varphi(k,\ell) = \frac{\operatorname{Im}[z(k)]\operatorname{Re}[G(k)] - \operatorname{Re}[z(k)]\operatorname{Im}[G(k)]}{k\operatorname{Re}[z(k)]}, \quad k \in \mathbf{R}, \tag{4.33}$$

$$\varphi'(k,\ell) = \frac{(r'_\ell\operatorname{Im}[z(k)] - k\,r_\ell)\operatorname{Re}[G(k)] - r'_\ell\operatorname{Re}[z(k)]\operatorname{Im}[G(k)]}{k\,r_\ell\operatorname{Re}[z(k)]}, \quad k \in \mathbf{R}. \tag{4.34}$$

From Proposition 2.1(d) we know that $\text{Re}[z(k)]$ is nonzero for $k \in \mathbf{R} \setminus \{0\}$. Using (4.33) and (4.34) in (4.32) we construct the Jost function $F(k)$. As stated in Proposition 3.1(c) the quantities $\varphi(k, \ell)$ and $\varphi'(k, \ell)$ are entire in $k \in \mathbf{C}$, and hence from (4.32) we see that $F(k)$ is also entire. Viewing $q(x)$ given in (3.3) as a compactly supported potential in $\mathbf{R}^+$, we can construct $q(x)$ for $x \in (0, \ell)$ from $F(k)$ known for $k \in \mathbf{R}$ via the Gel'fand-Levitan method as follows. As already mentioned, by Theorem 3.1 of [5] we know that $F(k)$ has one simple zero on the positive imaginary axis if and only if $r'_\ell < 0$ and that $F(k)$ is nonzero on the positive imaginary axis if and only if $r'_\ell \geq 0$. Because $F(k)$ is entire, its only possible zero on the positive imaginary axis is uniquely determined and we use $k = i\kappa_1$ to denote that zero, where $\kappa_1$ is some positive number. Such a zero corresponds to a bound state of the corresponding Schrödinger operator, and because of the compact support of the potential in $x \in \mathbf{R}^+$, the Gel'fand-Levitan norming constant $g_1$ corresponding to $k = i\kappa_1$ is determined from $F(k)$ as in (4.18) of [5] as

$$g_1^2 = \frac{-4i\kappa_1^2}{F(-i\kappa_1)\,\dot{F}(i\kappa_1)}, \tag{4.35}$$

where we recall that an overdot denotes the $k$-derivative. The potential $q(x)$ is constructed via the Gel'fand-Levitan method as follows [6,13,15,16]. We first construct the Gel'fand-Levitan kernel $G(x, y)$ as

$$G(x, y) := \frac{1}{\pi}\int_{-\infty}^\infty dk \left(\frac{k^2}{|F(k)|^2} - 1\right) \cos kx \cos ky + g_1^2 \cosh \kappa_1 x \cosh \kappa_1 y, \tag{4.36}$$

where it is understood that the second term on the right-hand side is absent when $r'_\ell \geq 0$. We use $G(x, y)$ as input to the Gel'fand-Levitan integral equation

$$A(x, y) + G(x, y) + \int_0^x ds\, A(x, s)\, G(s, y) = 0, \qquad 0 \leq y < x. \tag{4.37}$$

The potential $q(x)$ is recovered from the solution $A(x, y)$ to (4.37) as

$$q(x) = 2\frac{dA(x, x)}{dx}, \qquad x \in (0, \ell), \tag{4.38}$$

where $A(x, x)$ is understood to be $A(x, x^-)$. The regular solution $\varphi(k, x)$ is constructed from $A(x, y)$ as

$$\varphi(k, x) = \cos kx + \int_0^x dy\, A(x, y) \cos ky, \tag{4.39}$$

and we also have $r_0'/r_0 = A(0,0)$. ∎

In the proof of Proposition 4.5, using $\ell$ and $G(k)$ as input, we have described the recovery of the potential $q(x)$ and the regular solution $\varphi(k,x)$, respectively, via (4.38) and (4.39), which is based on using the Gel'fand-Levitan method. Let us remark that those two quantities can also be obtained by using various inverse spectral methods described in [19]. Having determined $\varphi(k,\ell)$ and $\varphi'(k,\ell)$ we also know their zeros. Thus, we can consider the Sturm-Liouville problem consisting the Schrödinger equation (3.2) and the two sets of boundary conditions given by

$$\begin{cases} \varphi'(k,0) - \dfrac{r_0'}{r_0}\varphi(k,0) = 0, & \varphi(k,\ell) = 0, \\ \varphi'(k,0) - \dfrac{r_0'}{r_0}\varphi(k,0) = 0, & \varphi'(k,\ell) = 0. \end{cases}$$

Then, Borg's uniqueness theorem [8,9] becomes applicable, and the quantities $q(x)$ and $\varphi(k,x)$ can be constructed by using various techniques presented in [19].

## 5. THE SOLUTION TO THE DIRECT PROBLEM

In Section 3 we have established the relevant properties of certain particular solutions to the Schrödinger equation and the relevant properties of the quantity $G(k)$ appearing in (3.15) and (3.18). Those properties will help us to obtain a useful representation of the solution $P(k,x)$ to the Webster horn equation (1.6) with the boundary condition (1.10) at $x=0$ and the boundary condition (1.14) at $x=\ell$, where $z(k)$ is the normalized impedance given in (1.15).

**Theorem 5.1** *Assume that the vocal-tract radius $r(x)$ belongs to the class $\mathcal{A}$ specified in Definition 1.1. Then, for $k \in \mathbf{C}$ the solution $P(k,x)$ to the Webster horn equation (1.6) satisfying the boundary conditions (1.10) and (1.14) exists, is unique, and is given by*

$$P(k,x) = -\frac{ikc\mu}{\pi r_0 \, r(x)} \left[ \frac{g(k,0)}{G(k)} \varphi(k,x) + S(k,x) \right], \qquad x \in [0,\ell], \tag{5.1}$$

*where $S(k,x)$ and $g(k,x)$ are the particular solutions to the Schrödinger equation (3.2) satisfying the respective initial conditions (3.5) and (3.7), $G(k)$ is the quantity appearing in (3.15) and (3.18), and $r_0$ is the positive constant appearing in (1.1).*

PROOF: From (3.1) we know that the general solution to (1.6) can be expressed in terms of two linearly independent solutions to (3.2). From (3.14) we see that $\varphi(k,x)$ and $S(k,x)$ are two linearly independent solutions to (3.2) for any $k \in \mathbf{C}$. Thus, we have

$$P(k,x) = \frac{1}{r(x)} \left[ \alpha(k)\,\varphi(k,x) + \beta(k)\,S(k,x) \right], \qquad x \in [0,\ell], \tag{5.2}$$

where $\alpha(k)$ and $\beta(k)$ are to be determined by requiring that the right-hand side of (5.2) satisfies (1.10) and (1.14). By taking the $x$-derivative of both sides of (5.2), we have

$$\begin{aligned}P'(k,x) =\ & -\frac{r'(x)}{r(x)^2} \left[ \alpha(k)\,\varphi(k,x) + \beta(k)\,S(k,x) \right] \\ & + \frac{1}{r(x)} \left[ \alpha(k)\,\varphi'(k,x) + \beta(k)\,S'(k,x) \right].\end{aligned} \tag{5.3}$$

Evaluating both sides of (5.3) at $x = 0$, and using (1.10), (3.5), and (3.6) in the resulting equation we get

$$-\frac{ikc\mu}{\pi r_0^2} = \frac{\beta(k)}{r_0},$$

from which we obtain

$$\beta(k) = -\frac{ikc\mu}{\pi r_0}. \tag{5.4}$$

Using (5.2) and (5.3) in (1.14), we obtain

$$\begin{aligned}& \left[ -\frac{r'_\ell}{r_\ell^2}\,\varphi(k,\ell) + \frac{1}{r_\ell}\,\varphi'(k,\ell) + \frac{ik}{z(k)\,r_\ell}\,\varphi(k,\ell) \right] \alpha(k) \\ & = -\left[ -\frac{r'_\ell}{r_\ell^2}\,S(k,\ell) + \frac{1}{r_\ell}\,S'(k,\ell) + \frac{ik}{z(k)\,r_\ell}\,S(k,\ell) \right] \beta(k).\end{aligned} \tag{5.5}$$

Since $r_\ell$ is positive and $z(k)$ is nonzero for $k \in \mathbf{R} \setminus \{0\}$, from (5.5) we get

$$\begin{aligned}& \left[ z(k)\,\varphi'(k,\ell) - \left( \frac{r'_\ell}{r_\ell}\,z(k) - ik \right) \varphi(k,\ell) \right] \alpha(k) \\ & = -\left[ z(k)\,S'(k,\ell) - \left( \frac{r'_\ell}{r_\ell}\,z(k) - ik \right) S(k,\ell) \right] \beta(k),\end{aligned} \tag{5.6}$$

which also holds at $k = 0$ as a result of Proposition 3.1(c) and the analyticity of $z(k)$ at $k = 0$. Using (3.7) in (5.6), we write the resulting equality in terms of the Wronskians as

$$[g(k,x); \varphi(k,x)]\,\alpha(k) = -[g(k,x); S(k,x)]\,\beta(k). \tag{5.7}$$

From (3.15) we see that the Wronskian on the left-hand side of (5.7) is equal to $-G(k)$. Since the Wronskian on the right-hand side of (5.7) is independent of $x$, it can be evaluated at $x=0$ and with the help of (3.5) we see that the aforementioned Wronskian is equal to $g(k,0)$. Thus, (5.7) is equivalent to

$$-G(k)\,\alpha(k) = -g(k,0)\,\beta(k). \tag{5.8}$$

Hence, from (5.4) and (5.8) we get

$$\alpha(k) = \frac{g(k,0)}{G(k)}\,\beta(k) = -\frac{ikc\mu}{\pi r_0}\,\frac{g(k,0)}{G(k)},$$

which establishes (5.1). We remark that by Proposition 3.1(c) the quantities $g(k,0)$, $\varphi(k,x)$, and $S(k,x)$ are entire in $k$, and by Proposition 3.2(a) the quantity $G(k)$ is entire. Hence, (5.1) is valid for $k \in \mathbf{C}$ and the poles of $P(k,x)$ in $k \in \mathbf{C}$ can only occur at the corresponding zeros of $G(k)$. ∎

For the analysis of the direct and inverse problems for (1.6) we need an expression for the pressure at the lips. That expression is given in the next proposition.

**Proposition 5.2** *Assume that the vocal-tract radius $r(x)$ belongs to the class $\mathcal{A}$ specified in Definition 1.1. Then, the quantity $P(k,\ell)$ representing the pressure at the lips is given by*

$$P(k,\ell) = -\frac{ikc\mu}{\pi r_0\,r_\ell}\,\frac{z(k)}{G(k)}, \qquad k \in \mathbf{C}, \tag{5.9}$$

*where $z(k)$ is the normalized impedance appearing in (1.14) and (1.15), $G(k)$ is the quantity appearing in (3.15) and (3.18), and $r_0$ and $r_\ell$ are the positive constants appearing in (1.1).*

PROOF: Evaluating (5.1) at $x=\ell$ we get

$$P(k,\ell) = -\frac{ikc\mu}{\pi r_0\,r_\ell}\,\frac{\Delta(k)}{G(k)}, \qquad k \in \mathbf{C}. \tag{5.10}$$

where we have defined

$$\Delta(k) := g(k,0)\,\varphi(k,\ell) + G(k)\,S(k,\ell). \tag{5.11}$$

From (3.5) and the definition of the Wronskian given in (3.11), we see that

$$g(k,0) = [g(k,x); S(k,x)], \tag{5.12}$$

and since the Wronskian in (5.12) is independent of $x$, it can be evaluated at $x = \ell$. Thus, from (5.12) we get

$$g(k,0) = g(k,\ell)\, S'(k,\ell) - g'(k,\ell)\, S(k,\ell). \tag{5.13}$$

On the other hand, from (3.15) with the right-hand side evaluated at $x = \ell$ we obtain

$$G(k) = \varphi(k,\ell)\, g'(k,\ell) - \varphi'(k,\ell)\, g(k,\ell). \tag{5.14}$$

Using (5.13) and (5.14) in (5.11) we obtain

$$\Delta(k) = g(k,\ell)\, [\varphi(k,\ell)\, S'(k,\ell) - \varphi'(k,\ell)\, S(k,\ell)], \tag{5.15}$$

where the quantity in the brackets is a Wronskian evaluated at $x = \ell$. However, that Wronskian is independent of $x$ and as seen from (3.14) the value of that Wronskian is equal to one. Thus, from (5.15) we obtain

$$\Delta(k) = g(k,\ell). \tag{5.16}$$

On the other hand, from the first equality in (3.7) we see that the right-hand side in (5.16) is equal to $z(k)$. Thus, (5.10) yields (5.9). ∎

From (1.5) we see that the volume velocity $V(k,x)$ can be expressed in terms of $P'(k,x)$ and that $V'(k,x)$ can be expressed in terms of $P(k,x)$. Thus, using the $x$-derivative of (5.1) in the first line of (1.5) we obtain the following result.

**Corollary 5.3** *Assume that the vocal-tract radius $r(x)$ belongs to the class $\mathcal{A}$ specified in Definition 1.1. Then, for $k \in \mathbf{C}$ the solution $P(k,x)$ and $V(k,x)$ to (1.5) satisfying the boundary conditions (1.10) and (1.14) exist and are unique. The quantity $P(k,x)$ is given by (5.1) and the quantity $V(k,x)$ is given by*

$$V(k,x) = \frac{r(x)}{r_0} \left[ \frac{g(k,0)}{G(k)} \varphi'(k,x) + S'(k,x) \right]$$
$$- \frac{r'(x)}{r_0} \left[ \frac{g(k,0)}{G(k)} + S(k,x)\varphi(k,x) \right], \qquad x \in [0,\ell],$$

where $S(k,x)$ and $g(k,x)$ are the particular solutions to the Schrödinger equation (3.2) satisfying the respective initial conditions (3.5) and (3.7), $G(k)$ is the quantity appearing in (3.15) and (3.18), and $r_0$ is the positive constant appearing in (1.1). We remark that $V(k,x)$ given in (5.1) is also the unique solution to (1.7) satisfying the boundary condition (1.9) at $x=0$ and the boundary condition, which is equivalent of (1.14), at $x=\ell$ expressed as

$$V'(k,\ell) + ik\, z(k)\, V(k,\ell) = 0.$$

## 6. THE INVERSE PROBLEM

In this section we consider the inverse problem of recovery of the vocal tract radius $r(x)$ in the class $\mathcal{A}$ specified in Definition 1.1 from an appropriate input data set related to some measurements taken at the lips. In particular we consider some input data sets related to the pressure $P(k,\ell)$ for $k \in \mathbf{R}^+$, the absolute pressure $|P(k,\ell)|$ for $k \in \mathbf{R}^+$, and the poles of $P(k,\ell)$ in $k \in \mathbf{C}$. We also indicate whether the knowledge of $\ell$ should be included in our input data set or else $\ell$ can be obtained from the available input data set used for the recovery.

As indicated in the proof of Proposition 4.5, the Jost function $F(k)$ appearing in (4.30) and (4.32) plays a key role as input in order to solve a relevant inverse problem to recover the potential $q(x)$ by the Gel'fand-Levitan method. Another relevant key quantity is given by $G(k)$ appearing in (3.15) and (3.18). We already know that the pressure $P(k,\ell)$ at the lips is related to $G(k)$ as in (5.9). Our main strategy to recover $r(x)$ from some input data set related to $P(k,\ell)$ is as follows. From the available input data set, we first construct $G(k)$ and then construct $F(k)$. We then use the Gel'fand-Levitan method to recover $r(x)$. The advantage of using the Gel'fand-Levitan method in the recovery of $r(x)$ is due to the fact that $r(x)$ is readily constructed from the regular solution $\varphi(k,x)$ as in the first equality in (3.11) and that $\varphi(k,x)$ itself is readily constructed as in (4.39) from the solution to the Gel'fand-Levitan equation (4.37).

In the next theorem we elaborate on (5.9), which describes the relationship among various key quantities.

**Theorem 6.1** *Assume that the vocal-tract radius $r(x)$ belongs to the class $\mathcal{A}$ specified in Definition 1.1. Let $P(k,x)$ be the corresponding pressure satisfying the boundary conditions (1.10) and (1.14). We then have the following:*

(a) *The knowledge of $|P(k,\ell)|$ for $k \in \mathbf{R}^+$ yields $|P(k,\ell)|$ for $k \in \mathbf{R}$, the constants $r_0$ and $r_\ell$, the normalized impedance $z(k)$, and the quantity $|G(k)|$ for $k \in \mathbf{R}$.*

(b) *The quantity $P(k,\ell)$ has a simple zero at $k=0$ and is meromorphic in $\mathbf{C}$ with its poles in $\mathbf{C} \setminus \{0\}$ coinciding with the nonzero zeros of $G(k)$ in $\mathbf{C}$. Hence, the set $\{k_j^-\}_{j=1}^{N^-}$ appearing in (4.1) corresponds to the poles of $P(k,\ell)$ in the fourth quadrant in $\mathbf{C}$, and the set $\{k_j^+\}_{j=1}^{N^+}$ appearing in (4.2) corresponds to the poles of $P(k,\ell)$ in the first quadrant.*

(c) *In addition to the yielded quantities mentioned in (a), the combined knowledge of $|P(k,\ell)|$ for $k \in \mathbf{R}^+$ and the poles (including multiplicities) of $P(k,\ell)$ in the fourth quadrant of $\mathbf{C}$ yields $G(k)$ for $k \in \mathbf{C}$, the quantity $E(k)$ for $k \in \mathbf{C}$, and the constant $\ell$.*

(d) *The knowledge of $P(k,\ell)$ for $k \in \mathbf{R}^+$ yields $P(k,\ell)$ for $k \in \mathbf{C}$ and hence also yields $G(k)$ for $k \in \mathbf{C}$, the quantity $E(k)$ for $k \in \mathbf{C}$, and the constant $\ell$.*

PROOF: The proof of (a) can be given as follows. From the first equality in (1.4) we see that $|P(-k,\ell)| = |P(k,\ell)|$ for $k \in \mathbf{R}$, showing that we know $|P(k,\ell)|$ for $k \in \mathbf{R}$ when we have $|P(k,\ell)|$ for $k \in \mathbf{R}^+$. From (5.9) we have

$$|P(k,\ell)| = \frac{|k|\,c\mu}{\pi r_0 r_\ell} \frac{|z(k)|}{|G(k)|}, \qquad k \in \mathbf{R}. \tag{6.1}$$

Letting $k \to 0$ in (6.1) and using (2.2) and (3.22) we obtain

$$\frac{|P(k,\ell)|}{|k|} = \frac{8c\mu}{3\pi^2 r_\ell} + O(k), \qquad k \to 0 \text{ in } \mathbf{R}. \tag{6.2}$$

Thus, $r_\ell$ is determined. On the other hand, letting $k \to \pm\infty$ in (6.1) and using (2.13) and (3.47) we get

$$|P(k,\ell)| = \frac{c\mu}{\pi r_0 r_\ell} + O\left(\frac{1}{\sqrt{|k|}}\right), \qquad k \to \pm\infty.$$

Hence, $r_0$ is also determined. Knowing $r_\ell$ we see from (1.15) that $z(k)$ is determined. Then, (6.1) reveals that $|G(k)|$ for $k \in \mathbf{R}$ is also determined. Thus, the proof of (a) is complete. Let us now prove (b). By Proposition 2.1(a) and Proposition 3.2(a), respectively, we know that $z(k)$ and $G(k)$ are entire. From (6.2) we know that $P(k,\ell)$ has a simple zero at $k=0$. Thus, using Proposition 3.2(e) in (5.9) we conclude that the zeros of $G(k)$ in $\mathbf{C} \setminus \{0\}$ correspond to the poles of $P(k,\ell)$. Hence, the proof of (b) is complete. Let us turn to the proof of (c). By (b) we know that the poles of $P(k,\ell)$ and the zeros of $G(k)$ in the fourth quadrant are equivalent and given by the set $\{k_j^-\}_{j=1}^{N^-}$ appearing in (4.1). Then, using (4.26) and (4.29) we can construct $e^{-ik\ell}G(k)$ for $k \in \overline{\mathbf{C}^-}$. Since $G(k)$ is entire, it then follows that $e^{-ik\ell}G(k)$ for $k \in \mathbf{C}$ and all the zeros of $G(k)$ in $\mathbf{C}$ are determined. Hence, we know $E(k)$ defined in (4.13). Then, as seen from (4.22) we also know $r_\ell + \ell$. However, from (a) we already know $r_\ell$. Thus, we also know the value of $\ell$. Thus, the proof of (c) is complete. For the proof of (d), we remark that (b) implies that the knowledge of $P(k,\ell)$ for $k \in \mathbf{R}^+$ yields $P(k,\ell)$ in $k \in \mathbf{C}$ including its poles. Thus, (c) applies and the proof of (d) is complete. ∎

In the next theorem we indicate that the combined knowledge of $\ell$ and the zeros of $P(k,\ell)$ on the right-half complex plane yields all the relevant information to determine $r(x)$.

**Theorem 6.2** *Assume that the vocal-tract radius $r(x)$ belongs to the class $\mathcal{A}$ specified in Definition 1.1. Let $P(k,x)$ be the corresponding pressure satisfying the boundary conditions (1.10) and (1.14). We then have the following:*

(a) *The combined knowledge of $\ell$ and the zeros of $P(k,\ell)$ in the right-half complex plane yields the quantity $E(k)$ in (4.13), the constants $r_\ell$ and $r_0$, the normalized impedance $z(k)$ for $k \in \mathbf{C}$, the quantity $G(k)$ for $k \in \mathbf{C}$, and the Jost function $F(k)$ for $k \in \mathbf{C}$ appearing in (4.30) and (4.32).*

(b) *The combined knowledge of $\ell$ and the zeros of $P(k,\ell)$ on the right-half complex plane is equivalent to the combined knowledge of $r_\ell$ and the zeros of $P(k,\ell)$ on the right-half* *complex plane. Hence, the latter set also yields the quantities mentioned in (a).*

PROOF: By Theorem 6.1(b) we know that the poles of $P(k, \ell)$ in $\mathbf{C}$ and the nonzero zeros of $G(k)$ in $\mathbf{C}$ coincide. Using (4.9)-(4.11) we conclude that the poles of $P(k, \ell)$ on the right-half complex plane determine $E(k)$ defined in (4.9). Then, using (4.22) we obtain $r_\ell$ and then we get $r_0$ via (4.24). Using $r_\ell$ in (1.15) we obtain $z(k)$. From (4.20) we obtain the constant $C$ and hence we also get $G(k)$ via (4.15). Next, using (3.45) we determine the two constants

$$-\gamma + \frac{r'_\ell}{r_\ell}, \quad -\gamma + \frac{2}{\pi r_\ell}, \tag{6.3}$$

where $\gamma$ is the constant defined in (3.46). Since we already have $r_\ell$, we recover $r'_\ell$ from the two constants listed in (6.3). Then, via (4.33) and (4.34) we get the quantities $\varphi(k, \ell)$ and $\varphi'(k, \ell)$. Finally, using (4.32) we obtain the Jost function $F(k)$. Since $z(k)$ and $G(k)$ are entire, their knowledge for $k \in \mathbf{R}$ is equivalent to their knowledge for $k \in \mathbf{C}$. Thus, the proof of (a) is complete. We remark that (b) directly follows from (4.22). ∎

In order to state our results in a concise and precise manner, we introduce several input data sets denoted by $\mathcal{D}_j$ for $1 \leq j \leq 6$. The following proposition shows that these six data sets are equivalent.

**Proposition 6.3** *Assume that the vocal-tract radius $r(x)$ belongs to the class $\mathcal{A}$ specified in Definition 1.1. Let $P(k, x)$ be the corresponding pressure satisfying the boundary conditions (1.10) and (1.14). Let $\{k_j^+\}_{j=1}^{N^+}$ and $\{k_j^-\}_{j=1}^{N^-}$ be the set of poles of $P(k, \ell)$ in the first and fourth quadrants in $\mathbf{C}$, respectively. Let*

$$\mathcal{D}_1 := \{P(k, \ell) : k \in \mathbf{R}^+\}, \tag{6.4}$$

$$\mathcal{D}_2 := \{|P(k, \ell)| : k \in \mathbf{R}^+; \{k_j^-\}_{j=1}^{N^-}\}, \tag{6.5}$$

$$\mathcal{D}_3 := \{\ell; \{k_j^-\}_{j=1}^{N^-}; \{k_j^+\}_{j=1}^{N^+}\}, \quad \mathcal{D}_4 := \{r_\ell; \{k_j^-\}_{j=1}^{N^-}; \{k_j^+\}_{j=1}^{N^+}\}, \tag{6.6}$$

$$\mathcal{D}_5 := \{\ell; E(k) : k \in \mathbf{C}\}, \quad \mathcal{D}_6 := \{r_\ell; E(k) : k \in \mathbf{C}\}. \tag{6.7}$$

*Then, these six data sets are equivalent.*

PROOF: We remark that Theorem 6.2(b) implies that $\mathcal{D}_3$ and $\mathcal{D}_4$ are equivalent and that $\mathcal{D}_5$ and $\mathcal{D}_6$ are equivalent. Because of Theorem 6.1(b) and the first equality in (1.4) we have the equivalence of $\mathcal{D}_3$ and $\mathcal{D}_5$. Using (a) and (b) of Theorem 6.1 we get the equivalence of $\mathcal{D}_1$ and $\mathcal{D}_2$. From Theorem 6.1(d) it follows that $\mathcal{D}_1$ contains $\mathcal{D}_5$ as a subset, and Theorem 6.2(a) implies that $\mathcal{D}_5$ contains $\mathcal{D}_1$ as a subset. This completes the proof that all the six input data sets appearing in (6.4)-(6.7) are equivalent. ∎

In the next theorem we show that any one of the six data sets described in (6.4)-(6.7), or any data set equivalent to one of those six data sets, uniquely determines the vocal tract radius.

**Theorem 6.4** *Assume that the vocal-tract radius $r(x)$ belongs to the class $\mathcal{A}$ specified in Definition 1.1. Let $P(k, x)$ be the corresponding pressure satisfying the boundary conditions (1.10) and (1.14). Any one of the input data sets described in (6.4)-(6.7), or any data set equivalent to one of those six data sets, uniquely determines $r(x)$.*

PROOF: By Theorem 6.2(a) we know that the Jost function $F(k)$ is determined for $k \in \mathbf{R}$. We can view $q(x)$ given in (3.3) as a real, integrable, compactly supported potential on the half line. Then, considering the Schrödinger equation (3.10) with the boundary condition at $x = 0$ described by the first equality in (3.30) we know [7] that the corresponding Jost function $F(k)$ is entire, it has either a simple zero at $k = 0$ or does not vanish at $k = 0$, its zeros in $\overline{\mathbf{C}^+} \setminus \{0\}$ can only occur on the positive imaginary axis and such zeros are simple and their number is finite. Then, we can use the Gel'fand-Levitan procedure summarized in (4.35)-(4.39) by using $F(k)$ as input into (4.36) and recover the regular solution $\varphi(k, x)$ as in (4.39). Hence, in particular we have $\varphi(0, x)$ for $x \in (0, \ell)$. As indicated in Theorem 6.2(a), we also have $r_0$. Then, we use the first equality in (3.11) to recover $r(x)$ as

$$r(x) = r_0 \, \varphi(0, x), \qquad x \in (0, \ell). \tag{6.8}$$

Thus, the proof is complete. ∎

Even though the construction of $r(x)$ is outlined in the proofs of Theorems 6.1, 6.2, and 6.4, for the convenience of the reader we provide an orderly summary of those steps below. The steps are given for the input data set $\mathcal{D}_5$ defined in (6.7) and can easily be adapted for other equivalent input data sets. Hence, the following are the steps to construct $r(x)$ when $\ell$ and the quantity $E(k)$ defined in (4.9) are known.

(a) Using (4.20), obtain the constant $C$.

(b) Using (4.22), determine $r_\ell$.

(c) Using $r_\ell$ in (1.15), obtain $z(k)$.

(d) Using $r_\ell$ in (4.24), obtain $r_0$.

(e) Using (4.15), determine $G(k)$ for $k \in \mathbf{C}$.

(f) Using (3.45), with the help of (6.3), obtain $r'_\ell$.

(g) Using (4.33) and (4.34), determine $\varphi(k, \ell)$ and $\varphi'(k, \ell)$, respectively.

(h) Using (4.32), obtain $F(k)$.

(i) If $r'_\ell < 0$, determine the only zero of $F(k)$ on the positive imaginary axis occurring at $k = i\kappa_1$, where $\kappa_1$ is the constant appearing in (4.36). If $r'_\ell \geq 0$, then this step can be omitted.

(j) If $r'_\ell < 0$, then using (4.35) determine $g_1^2$. If $r'_\ell \geq 0$, then this step can be omitted and we recall that the term in (4.36) containing $\kappa_1$ is missing from the definition of $G(x, y)$.

(k) Use (4.36) as input to the Gel'fand-Levitan integral equation (4.37).

(l) By solving (4.37), obtain $A(x, y)$ for $0 \leq y < x \leq \ell$.

(m) Using (4.39), determine $\varphi(k, x)$ for $x \in (0, \ell)$.

(n) Recover $r(x)$ by using (6.8).

## 7. EXAMPLES

In this section we present some explicit examples to illustrate the theory presented in the previous sections.

We recall that the key quantity $G(k)$ appearing in (3.15) and (3.18) is closely related to the pressure at the lips, and as seen from (5.9) the zeros of $G(k)$ in $\mathbf{C}$ corresponds to the poles of $P(k,\ell)$. We can determine $G(k)$ in terms of the Jost solution $f(k,x)$ appearing in (3.4). Expressing $g(k,x)$ appearing in (3.7) as a linear combination of $f(k,x)$ and $f(-k,x)$, evaluating the resulting equation at $x = 0$ and $x = \ell$, respectively, and using (3.4) and (3.7), we obtain

$$\begin{bmatrix} g(k,0) \\ g'(k,0) \end{bmatrix} = \begin{bmatrix} f(k,0) & f(-k,0) \\ f'(k,0) & f'(-k,0) \end{bmatrix} \begin{bmatrix} e^{ik\ell} & e^{-ik\ell} \\ ik\, e^{ik\ell} & -ik\, e^{-ik\ell} \end{bmatrix}^{-1} \begin{bmatrix} z(k) \\ \dfrac{r'_\ell}{r_\ell} z(k) - ik \end{bmatrix}. \quad (7.1)$$

We can express $G(k)$ by evaluating the Wronskian on the right-hand side of (3.15) at $x = 0$, and we get

$$G(k) = \begin{bmatrix} -\varphi'(k,0) & \varphi(k,0) \end{bmatrix} \begin{bmatrix} g(k,0) \\ g'(k,0) \end{bmatrix}. \quad (7.2)$$

Using (3.6) and (7.1) on the right-hand side of (7.2) we have

$$G(k) = \begin{bmatrix} -\dfrac{r'_0}{r_0} & 1 \end{bmatrix} \begin{bmatrix} f(k,0) & f(-k,0) \\ f'(k,0) & f'(-k,0) \end{bmatrix} \begin{bmatrix} e^{ik\ell} & e^{-ik\ell} \\ ik\, e^{ik\ell} & -ik\, e^{-ik\ell} \end{bmatrix}^{-1} \begin{bmatrix} z(k) \\ \dfrac{r'_\ell}{r_\ell} z(k) - ik \end{bmatrix}. \quad (7.3)$$

If we know the Jost solution $f(k,x)$ explicitly, we can use (7.3) to obtain $G(k)$ explicitly.

In order to understand the properties of $G(k)$ better, let us consider some explicit examples. In the first example below, we consider a uniform tube of constant radius to describe the vocal tract.

**Example 7.1** Let us assume that the vocal tract radius $r(x)$ is equal to $b$ for some appropriate constant $b$. Let us use two free parameters given by $r_0$ and $\ell$. Thus, we have

$$b = r_0, \quad r_\ell = r_0, \quad r'_0 = r'_\ell = 0.$$

From (3.3) we see that $q(x) \equiv 0$, and hence the corresponding Jost solution appearing in (3.4) is given by $f(k,x) = e^{ikx}$. Then, from (7.3) we obtain

$$G(k) = -ik\cos k\ell + k\, z(k)\sin k\ell,$$

where we remark that the dependence on $r_\ell$ is through the normalized impedance $z(k)$. One can check the properties of $G(k)$ listed in the previous sections by using some input values such as $r_0 = 1$ and $\ell = 17$, where the centimeter is used as the length unit.

In the next example, we consider the vocal tract radius as a linear function of $x$ and evaluate the corresponding quantity $G(k)$ explicitly.

**Example 7.2** Let us consider the radius $r(x)$ as equal to $ax + b$ for some appropriate constants $a$ and $b$. We can use three free parameters given by $r_0$, $r'_0$, and $\ell$. We then have

$$a = r'_0, \quad b = r_0, \quad r_\ell = r_0 + \ell r'_0, \quad r'_\ell = r'_0.$$

In this case, from (3.3) we see that $q(x) \equiv 0$ and hence $f(k, x) = e^{ikx}$. Then, from (7.3) we get

$$G(k) = \frac{e^{-ik\ell} Q_1 + e^{ik\ell} Q_2}{2kr_0(r_0 + \ell r'_0)},$$

where we have defined

$$Q_1 := -(r'_0 - ikr_0)\left[k\ell r'_0(-1 + z) + kr_0(-1 + z) - iz\, r'_0\right],$$

$$Q_2 := -(r'_0 + ikr_0)\left[k\ell r'_0(1 + z) + kr_0(1 + z) + iz\, r'_0\right],$$

with $z$ denoting $z(k)$. For example, using some particular values for the free parameters such as

$$r_0 = \frac{1}{\sqrt{5}}, \quad r'_0 = \frac{2(\sqrt{5} - 1)}{85}, \quad \ell = 17,$$

$$r_0 = \frac{1}{\sqrt{5}}, \quad r'_0 = -\frac{\sqrt{5} - 1}{85}, \quad \ell = 17,$$

one can investigate various properties of $G(k)$ numerically.

In the next example, we consider the vocal tract radius $r(x)$ as a quadratic function of $x$.

**Example 7.3** Let us assume that the vocal tract radius $r(x)$ is equal to $(ax+b)^2$ for some appropriate constants $a$ and $b$. Let us use three free parameters given by $r_0$, $r'_0$, and $\ell$. We have

$$a = \frac{r'_0}{2\sqrt{r_0}}, \quad b = \sqrt{r_0}, \quad r_\ell = \left(\sqrt{r_0} + \frac{\ell r'_0}{2\sqrt{r_0}}\right)^2, \quad r'_\ell = r'_0 + \frac{\ell (r'_0)^2}{2r_0}.$$

In this case, the potential $q(x)$ is given by

$$q(x) = \frac{2a^2}{(ax+b)^2}, \qquad x \in (0, \ell),$$

and the Jost solution $f(k, x)$ can be constructed as a linear combination of the two solutions to (2.3) that are given by $e^{ikx} m(k, x)$ and $e^{-ikx} m(k, x)$, where

$$m(k, x) := 1 + \frac{ia}{k(ax+b)}, \tag{7.4}$$

in such a way that $f(k, x)$ satisfies (3.4). In this case, with the help (7.4) from (7.3) we obtain

$$G(k) = \frac{Q_1 \cos k\ell + Q_2 \sin k\ell}{8ik^3 r_0^2 (2r_0 + \ell r_0')^2}, \tag{7.5}$$

where $z$ denotes $z(k)$ and we have

$$Q_1 := -18ikz\ell(r_0')^4 - k^2 \left[ 12\ell r_0 (r_0')^3 + 6\ell^2 (r_0')^4 \right] + ik^3 Q_3 + k^4 Q_4,$$

$$Q_2 := 18iz(r_0')^4 + k \left[ 12 r_0 (r_0')^3 + 6\ell (r_0')^4 \right] + ik^2 Q_5 + k^3 Q_6 + ik^4 Q_7,$$

$$Q_3 := -24z\ell r_0^2 (r_0')^2 - 12z\ell^2 r_0 (r_0')^3, \quad Q_4 := 32 r_0^4 + 32\ell r_0^3 r_0' + 8\ell^2 r_0^2 (r_0')^2,$$

$$Q_5 := 24 z r_0^2 (r_0')^2 + 12 z \ell r_0 (r_0')^3 - 6 z \ell^2 (r_0')^4, \quad Q_6 := 32 r_0^3 r_0' + 40\ell r_0^2 (r_0')^2 + 12\ell^2 r_0 (r_0')^3,$$

$$Q_7 := 32 z r_0^4 + 32 z \ell r_0^3 r_0' + 8 z \ell^2 r_0^2 (r_0')^2.$$

Using the input values

$$r_0 = \frac{1}{\sqrt{5}}, \quad r_0' = \frac{2(\sqrt{5}-1)}{85}, \quad \ell = 17, \tag{7.6}$$

we get

$$r(x) = \left( \frac{5^{3/4} - 5^{1/4}}{85} x + \frac{1}{5^{1/4}} \right)^2, \qquad x \in (0, 17), \tag{7.7}$$

whose graph is indicated in Figure 7.1.

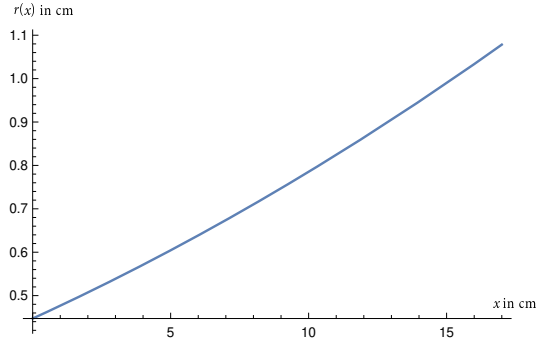

**Figure 7.1.** The vocal-tract radius $r(x)$ given in (7.7).

We can numerically investigate the corresponding $G(k)$ given in (7.5) with input from (7.6). The zeros of $G(k)$ in the first quadrant in $\mathbf{C}$ are shown in Figure 7.2.

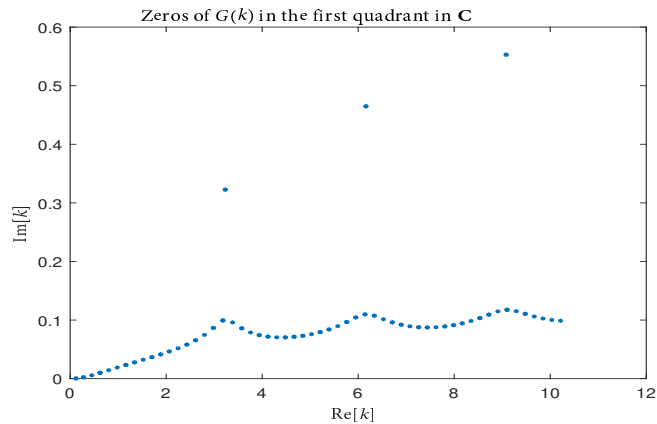

**Figure 7.2.** The location of zeros of $G(k)$ in the first quadrant with $r(x)$ as in (7.7).

From Figure 7.2 we observe that in the first quadrant of $\mathbf{C}$, the key quantity $G(k)$ has 11 zeros when $|k| < 2$, it has 28 zeros when $|k| < 5$, and it has 57 zeros when $|k| < 10$. Note that the density of the zeros of $G(k)$ in the first quadrant as $k \to \infty$ is predicted in Proposition 4.1(c) as $(r_\ell + \ell)/\pi$, which is $5.754\overline{5}$, with the overline an a digit indicating a round off. A numerical investigation can also be carried out by using another set of values such as

$$r_0 = \frac{1}{\sqrt{5}}, \quad r'_0 = -\frac{\sqrt{5}+1}{85}, \quad \ell = 17.$$

**Acknowledgments.** The third author is supported in part by the National Natural Science Foundation of China (119001304) and the Startup Foundation for Introducing Talent of NUIST.